\documentclass[12pt,preprint]{aastex}
\usepackage{emulateapj5}
\usepackage{apjfonts}
\usepackage{epsfig}
\usepackage{natbib}

\newcommand{\caii}{\ion{Ca}{2}}
\newcommand{\cat}{CaT}
\newcommand{\chk}{Ca~H+K}
\newcommand{\chisq}{\ensuremath{\chi^2}}

\newcommand{\etal}{et al.}

\newcommand{\feii}{\ion{Fe}{2}}

\def\gtrsim{\mathrel{\hbox{\rlap{\hbox{\lower4pt\hbox{$\sim$}}}\hbox{\raise2pt\hbox{$>$}}}}}

\newcommand{\fwhb}{\ensuremath{\mathrm{FWHM}_\mathrm{H{\beta}}}}
\newcommand{\halpha}{H\ensuremath{\alpha}}

\newcommand{\hbeta}{H\ensuremath{\beta}}

\newcommand{\kms}{km~s\ensuremath{^{-1}}}

\newcommand{\ldelsig}{\ensuremath{\delta \sigma_{\ast}}}

\newcommand{\lum}{ergs s$^{-1}$}

\newcommand{\lledd}{\ensuremath{L_{\mathrm{bol}}/L{\mathrm{_{Edd}}}}}

\newcommand{\mbh}{\ensuremath{M_\mathrm{BH}}}

\newcommand{\mgb}{\ion{Mg}{1}$b$}
\newcommand{\msigma}{\ensuremath{M_{\mathrm{BH}}-\sigmastar}}
\newcommand{\msun}{\ensuremath{M_{\odot}}}

\newcommand{\oii}{[\ion{O}{2}]}

\newcommand{\rblr}{\ensuremath{R_{\mathrm{BLR}}}}

\newcommand{\sigmastar}{\ensuremath{\sigma_{\ast}}}

\newcommand{\zw}{{\small I}~Zw~1}

\def\lax{{$\mathrel{\hbox{\rlap{\hbox{\lower4pt\hbox{$\sim$}}}\hbox{$<$}}}$}}
\def\gax{{$\mathrel{\hbox{\rlap{\hbox{\lower4pt\hbox{$\sim$}}}\hbox{$>$}}}$}}

\slugcomment{To appear in {\it The Astrophysical Journal}.}
\shorttitle{Measuring \sigmastar\ in AGNs}
\shortauthors{GREENE \& HO}

\begin{document}

\title{Measuring Stellar Velocity Dispersions in Active Galaxies}

\author{Jenny E. Greene}
\affil{Harvard-Smithsonian Center for Astrophysics, 60 Garden St., 
Cambridge, MA 02138}

\and

\author{Luis C. Ho}
\affil{The Observatories of the Carnegie Institution of Washington,
813 Santa Barbara St., Pasadena, CA 91101}

\begin{abstract}

We present stellar velocity dispersion (\sigmastar) measurements for a
significant sample of 40 broad-line (Type 1) active galaxies for use
in testing the well-known relation between black hole mass and stellar
velocity dispersion.  The objects are selected to contain Ca~{\sc II}
triplet, \mgb\ triplet, and Ca H+K stellar absorption features in
their optical spectra so that we may use them to perform extensive
tests of the systematic biases introduced by both template mismatch
and contamination from the active galactic nucleus (AGN).  We use the
Ca~{\sc II} triplet as a benchmark to evaluate the utility of the
other spectral regions in the presence of AGN contamination.  Broad
\feii\ emission, extending from $\sim 5050-5520$ \AA, in combination
with narrow coronal emission lines, can seriously bias \sigmastar\
measurements from the \mgb\ region, highlighting the need for extreme
caution in its use.  However, we argue that at luminosities
constituting a moderate fraction of the Eddington limit, when the
\feii\ lines are both weak and smooth relative to the stellar lines,
it is possible to derive meaningful measurements with careful
selection of the fitting region.  In particular, to avoid the
contamination of coronal lines, we advocate the use of the region
5250--5820 \AA, which is rich in Fe absorption features.  At higher AGN
contaminations, the \chk\ region may provide the only recourse for
estimating \sigmastar.  These features are notoriously unreliable, due
to a strong dependence on spectral type, a steep local continuum, and
large intrinsic broadening.  Indeed, we find a strong systematic trend
in comparisons of \chk\ with other spectral regions.  Luckily the
offset is well-described by a simple linear fit as a function of
\sigmastar, which enables us to remove the bias, and thus extract
unbiased \sigmastar\ measurements from this region.  We lay the
groundwork for an extensive comparison between black hole mass and
bulge velocity dispersion in active galaxies, as described in a
companion paper by Greene \& Ho.
\end{abstract}


\section{Introduction}

It has long been known that some galaxies harbor supermassive black holes 
(BHs) at their centers, whose accretion-powered luminosity may outshine their 
entire host galaxy (e.g.,~Lynden-Bell 1969).  More recently, we have learned 
that most, if not all, galaxies with a bulge contain central BHs (Kormendy \& 
Richstone 1995; Magorrian \etal\ 1998; Ho 1999), whose masses are tightly 
correlated with the stellar velocity dispersion 
(\sigmastar) of the bulge (the \msigma\ relation: Ferrarese \& Merritt 2000;
Gebhardt \etal\ 2000a; Tremaine \etal\ 2002).  It is likely that the
\msigma\ relation is established during the active galactic nucleus
(AGN) phase of a galaxy's life-cycle, since energy emitted by the BH may
simultaneously limit the gas supply for building both the bulge and the BH
itself (e.g.,~Silk \& Rees 1998).  If we are to understand this
critical stage in galaxy evolution, we must have robust methods to 
estimate both BH masses and \sigmastar\ in active galaxies.  

Reverberation mapping (Blandford \& McKee 1982) is currently the most
direct way to obtain BH masses in AGNs, although the masses are uncertain by 
a factor that depends on the poorly constrained geometry of the broad-line
region (BLR).  Reverberation mapping provides an estimate of the size
of the BLR from the lag between the variability in the photoionizing
continuum and the broad emission lines.  With a BLR size at hand, one
can infer a virial mass for the very central region of the AGN, and
hence for the BH (Ho 1999; Wandel \etal\ 1999; Kaspi et al. 2000;
Peterson et al. 2004): \mbh\ = $fR_{\mathrm{BLR}} v^2/G$, where $v$ is
the velocity dispersion of the BLR gas and the factor $f$ accounts for
the geometry of the BLR (e.g.,~Onken \etal\ 2004; Kaspi \etal\
2005).  For instance, a spherical BLR has an $f$ value of 0.75 when
$v$ is measured using the full width at half-maximum (FWHM) of the
broad line (e.g.,~Netzer 1990).  Reverberation mapping has further
been used to derive an empirical relation between AGN luminosity and
BLR radius (the radius-luminosity relation): \rblr\ $\propto L_{5100
\AA}^{0.6-0.7}$ (Kaspi \etal\ 2000, 2005; Greene \& Ho 2005b). Since
reverberation mapping data are time-consuming to obtain, and are
currently available only for a small number of objects, practical
applications of the virial technique to estimate BH masses have relied
on the radius-luminosity relation and BLR line widths measured from
single-epoch spectra (e.g.,~McLure \& Dunlop 2001; Vestergaard 2002;
Greene \& Ho 2004).  Given the many uncertainties inherent in the
virial technique (see Krolik 2001 for a detailed discussion), in
particular the unknown geometry of the BLR, one may justifiably question its 
reliability.  On the other hand, Gebhardt \etal\ (2000b) and Ferrarese
\etal\ (2001) showed that virial BH mass estimates agree surprisingly
well with masses inferred from the \msigma\ relation, at least for a
handful of objects.  Subsequent work (Nelson et al. 2004; Onken \etal\
2004) has increased the sample size, still considering a total of only
17 objects between the two samples.  Onken \etal\ find that
reverberation mapped masses, using the standard assumption of a
spherical BLR and the FWHM, are $\sim 0.26$ dex below the masses
inferred from the Tremaine \etal\ (2002) fit to the \msigma\ relation;
Nelson et al. obtain essentially the same result, concluding that the
AGN sample lie systematically below the sample of inactive galaxies by
0.21 dex.  However, as noted by Nelson \etal, the significance of the
measured offset is low, given the large scatter (0.46 dex) in the
data.  Since this value represents the zeropoint in the BH mass scale
for AGNs, many more \sigmastar\ measurements are required to improve
the calibration.  An enlarged sample would allow us to resolve
outstanding technical questions, such as whether the FWHM or the
actual second moment of the broad-line profile is a better measure of
BLR velocity dispersion (e.g.,~Vestergaard 2002; Peterson \etal\
2004), as yet unaddressed for virial masses obtained with single-epoch
spectra.  Furthermore, we would be able to explore the importance of
BLR geometry in a statistical way, both by addressing the possible
importance of inclination (e.g.,~McLure \& Dunlop 2001) and by
obtaining a more secure empirical measure of $f$.  The BLR geometry
may even depend on properties of the system such as BH mass and
accretion rate.  With such calibrations in hand, we may begin to
address other fundamental questions, such as whether the \msigma\
relation varies as a function of BH mass (e.g.,~Robertson \etal\ 2005)
or evolves with time (e.g.,~Shields \etal\ 2003).

The primary hurdle in this endeavor lies with the difficulty of obtaining 
reliable \sigmastar\ measurements in active galaxies.  AGNs are bright, and
thus detectable to cosmological distances, but the strong continuum of Type 1 
sources dilutes the starlight while their rich emission-line spectrum confuses 
and distorts the shape of the stellar absorption features.  This paper 
presents a comprehensive discussion of how to tackle this problem.  Using a 
relatively large sample of AGNs for which we can measure \sigmastar\ reliably, 
we discuss the relative merits and complications of measuring \sigmastar\ 
from different spectral regions.  The sample is described in \S2.  
Section 3 introduces the direct-fitting code we have developed for measuring 
\sigmastar\ and discusses the battery of tests to which we have subjected it 
to evaluate its robustness and limitations.   The many challenges inherent 
in dealing with AGN spectra are outlined in \S4, where we devote considerable
attention to finding the optimal spectral regions for measuring \sigmastar\ 
under realistic conditions encountered in AGNs.  We end with some practical 
suggestions to serve as a guide for other researchers (\S5), followed by a 
summary (\S6).  A companion paper (Greene \& Ho 2005c) uses the final 
measurements obtained from this analysis to provide a new appraisal of the 
\msigma\ relation of AGNs.

Throughout we assume the following cosmological parameters to calculate
distances: $H_0 = 100~h = 71$~\kms~Mpc$^{-1}$, $\Omega_{\rm m} = 0.27$,
and $\Omega_{\Lambda} = 0.75$ (Spergel \etal\ 2003).

\section{The Sample}

We utilize the large and homogeneous database provided by the Sloan
Digital Sky Survey (SDSS; York \etal\ 2000), which will eventually
obtain imaging and follow-up spectroscopy for about one-quarter of the
sky.  Briefly, spectroscopic candidates are selected based on
multi-band imaging (Strauss \etal\ 2002) with a drift-scan camera
(Gunn \etal\ 1998).  A pair of spectrographs is fed by
$3\arcsec$-diameter fibers, covering $\sim 3800 - 9200$ \AA\ with an
instrumental resolution of $\lambda/\Delta \lambda \approx 1800$
(Gaussian $\sigma_{\rm inst} \approx 70$ \kms).  The spectroscopic
pipeline performs basic image calibrations, as well as spectral
extraction, sky subtraction, removal of atmospheric absorption bands,
and wavelength and spectrophotometric calibration (Stoughton \etal\
%%%%%%%%%%%%%%%%%%%%%%%%%%%%%%%%%%%%%%%%%%%%%%%%%%%%%%%%%%%%%%%%%%%%
%%BoundingBox: 
\psfig{file=table1_v3.epsi,width=0.45\textwidth,keepaspectratio=true,angle=0}
\vskip 2mm
%%%%%%%%%%%%%%%%%%%%%%%%%%%%%%%%%%%%%%%%%%%%%%%%%%%%%%%%%%%%%%%%%%%%
\noindent
2002).  Finally, spectral classification is performed using
cross-calibration with stellar, emission-line, and active galaxy
templates. The Third Data Release (DR3; Abazajian \etal\ 2005) of SDSS
contains 45,260 spectroscopically identified AGNs with $z < 2.3$.  

%%%%%%%%%%%%%%%%%%%%%%%%%%%%%%%%%%%%%%%%%%%%%%%%%%%%%%%%%%%%%%%%%%%%
\begin{figure*}[t]
\begin{center}
\vskip +0.2truein
\vbox{
%\vskip -0.2truein
\hbox{
\hskip -0.1truein
\psfig{file=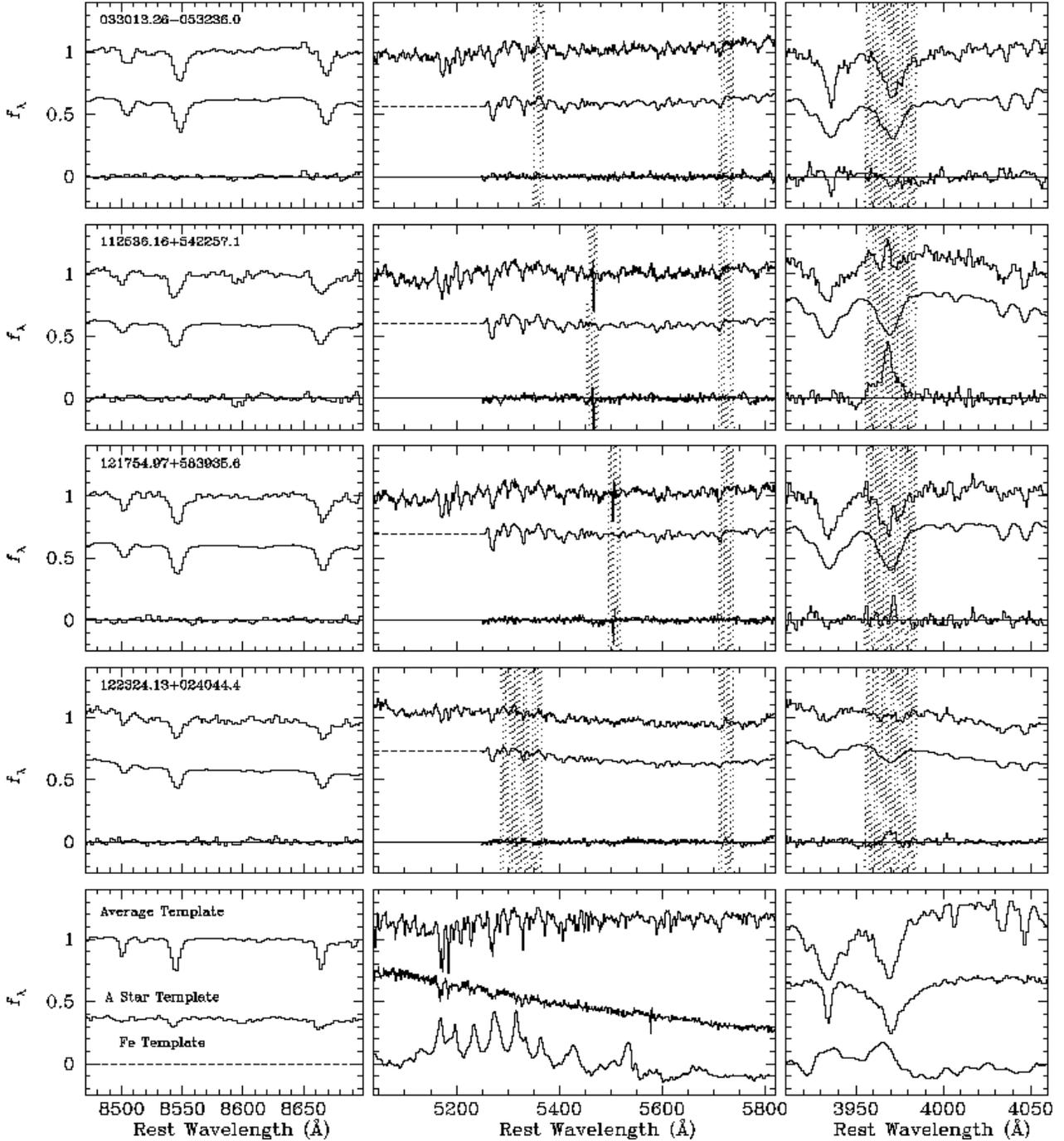,width=6.1in.,angle=0}
}}
\vskip +6mm 
\figcaption[]{ 
Sample spectral fits for the regions around ({\it left column}) the
Ca~{\tiny II} $\lambda \lambda$~8498, 8542, 8662 triplet, ({\it middle
column}) the Mg~{\tiny I}{\it b} $\lambda \lambda$5167, 5173, 5184
triplet, and ({\it right column}) Ca~H+K $\lambda \lambda$3969, 3934.
The {\it top four rows}\ are objects from our sample; the official
SDSS name is given in the upper left-hand corner for each row. All
spectra are shown at rest wavelengths and have been normalized to
their local continua.  The shaded areas denote regions that were
excluded from the fit.  For each object, the top spectrum is the
galaxy spectrum, the middle spectrum is our best-fit template model
(offset for clarity), and the bottom spectrum is the residuals.  Note
that the CaT features are much less diluted than Mg~{\tiny I}{\it b}.
In the Mg~{\tiny I}{\it b} region we have only fit the range
5250--5820 \AA, but we include the Mg~{\tiny I}{\it b} features to
specifically highlight their high level of AGN dilution and Fe~{\tiny
II} contamination.  In the Ca~H+K region the entire H line needs to be
masked from the fit.  The {\it bottom row}\ shows the templates used
in the analysis: the average stellar template, the A star template, and the Fe
template derived from I~Zw~1 (see text for details). The current Fe
template does not extend to the CaT region, but Fe~{\tiny II} emission
is known to be weak in this region for AGNs (e.g.,~Persson 1988).
\label{contfig}}
\end{center}
\end{figure*}
%%%%%%%%%%%%%%%%%%%%%%%%%%%%%%%%%%%%%%%%%%%%%%%%%%%%%%%%%%%%%%%%%%%%
%\noindent

There are a limited number of optical stellar spectral features that
have high enough equivalent width (EW) to be visible even in the 
presence of significant AGN contamination (see Fig. 1).  The most
common spectral regions employed are either around the \caii\ $\lambda
\lambda$8498, 8542, 8662 triplet (hereafter CaT) or the
\ion{Mg}{1}{\it b} $\lambda \lambda$5167, 5173, 5184 triplet.  Since
we plan to compare various spectral regions, we select all
spectroscopically identified AGNs from DR3 with $z \leq 0.05$, such
that at least two of the three CaT lines are in the band.  Of the 66
objects satisfying this criterion, we limit our attention to those
with \caii\ $\lambda 8542$ EW $\geq$ 1.5 \AA\ and a local
signal-to-noise ratio (S/N) of 18 or higher per pixel.  On inspection, we
rejected nine candidates meeting these requirements because of
confusion from either Paschen or \ion{O}{1}~$\lambda 8446$ emission
features or strong night-sky line residuals, leaving a final sample of
40 objects (Table 1).

\section{Stellar Velocity Dispersion Measurements}

\subsection{The Method}

Depending on the application, a variety of methods are used to measure stellar
velocity dispersions.  Those involving Fourier techniques
include cross-correlation (Tonry \& Davis 1979), the Fourier quotient
(Simkin 1974; Sargent et~al. 1977), and the Fourier correlation
quotient (Bender 1990).  The use of Fourier space is natural, as
velocity broadening is reduced to a multiplication, making the methods
computationally cheap. It is also possible, however, to compare a grid
of broadened template spectra directly with the galaxy spectrum in
pixel space (Burbidge et al. 1961).  While direct-fitting
is comparatively computationally expensive, it is not particularly
demanding by modern standards.  There are many advantages to
direct-fitting, as described, for example, by Rix \& White (1992).  Since
the quality of the fit is measured in pixel space, error analysis and the
effects of noise are easily incorporated, as is the masking of
corrupted spectral regions or emission lines. Furthermore, AGN
features are easily incorporated as additional model components in
a direct fit.  Following Barth \etal\ (2002a), we build a
model, $M(x)$, that is the convolution of a stellar template spectrum, $T(x)$,
and a line-of-sight velocity broadening function approximated as a Gaussian, 
$G(x)$:

\begin{equation}
M(x) = P(x)~\{[T(x)~\otimes\ G(x)] + C(x)\}.
\end{equation}

\noindent
The spectra and templates are normalized to the local continuum using
identical spectral regions free of strong emission lines.  We assume a
Gaussian velocity broadening function for simplicity, but technically
it is possible to solve for higher-order moments of the velocity
profile (e.g.,~Rix \& White 1992; van~der~Marel 1994). In our case the
data quality does not warrant such a treatment.  $C(x)$ is a model for
the AGN continuum, here assumed to be a single power law, 
where both the
normalization and slope are allowed to vary in the fit.  This
component could comprise other additive components, such as the
\feii\ emission from the BLR of the AGN that forms a ``pseudo-continuum''
throughout the optical spectrum (e.g., Francis et al. 1991).  The polynomial
factor, $P(x)$, is required to account for variations in continuum
shape between the template and the galaxy (see, e.g.,~Kelson \etal\
2000), which, in our case, can result from a combination of internal
reddening in the host galaxy (Galactic extinction is already accounted
for), differing stellar populations, and residual calibration errors.
We typically use a third-order Legendre polynomial.  Altogether, in
addition to the velocity dispersion, we solve for the velocity shift,
an amplitude and slope for the power-law component, and four
polynomial components, for a total of eight free parameters.  The
best-fit parameters are determined through a minimization of the
\chisq\ statistic using the non-linear Levenberg-Marquardt
minimization algorithm as implemented by {\it mpfit}\ in IDL.

For stellar templates, we use 32 G and K giant stars in the old open
cluster M67 observed with SDSS (see Bernardi \etal\ 2003).  While the
use of field stars is far more common, these templates are preferable
because they are observed in an identical manner as the program
objects.  The near-solar metallicity of M67 ([Fe/H] =
$-$0.05$\pm0.03$; Montgomery \etal\ 1993) means that template mismatch
problems resulting from metallicity (discussed below) are comparable
to those expected from field stars.  Note that we are using a
restricted range of stellar spectral types for our templates, while a
wider range of stellar types must be present at some level in the
galaxies. If we had sufficient S/N, and AGN contamination were not serious,
we could determine the optimal combination of template stars to 

%%%%%%%%%%%%%%%%%%%%%%%%%%%%%%%%%%%%%%%%%%%%%%%%%%%%%%%%%%%%%%%%%%%%
%%BoundingBox: 50 180 540 720
\hskip -0.3in
\psfig{file=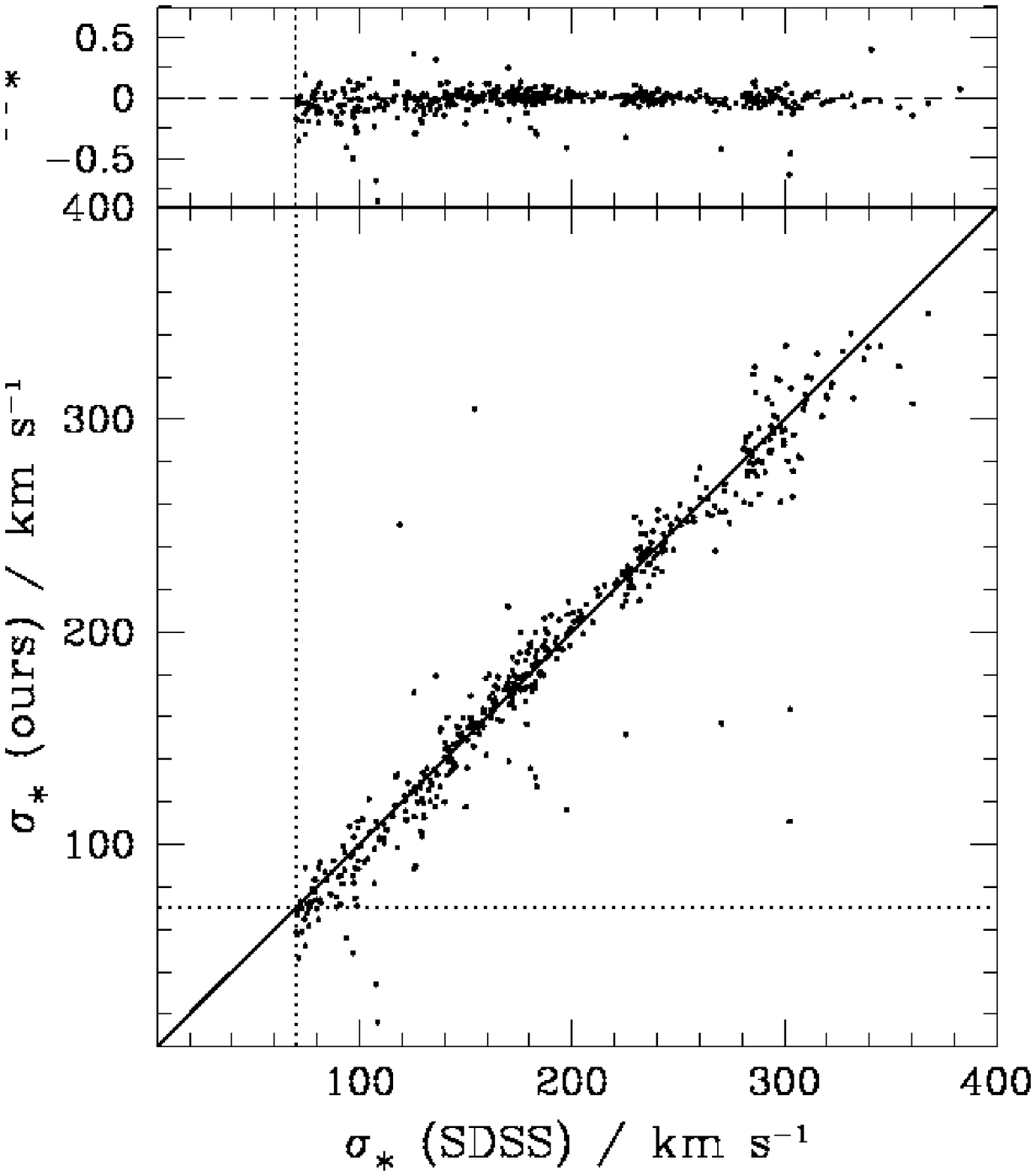,width=0.52\textwidth,keepaspectratio=true,angle=0}
\vskip -0mm
\figcaption[]{
Comparison between $\sigmastar$ measurements from SDSS (Heckman \etal\ 2004)
and our own measurements of a select sample of
504 galaxies from that work.  The agreement is reasonable, since the
discrepancies at the low and high end correspond to regions with lower
S/N spectra.  The solid line represents equality of the two
measurements, and the dotted lines mark the nominal resolution
limit of the SDSS.  The top panel shows the fractional difference
\ldelsig $\equiv$ [$\sigma{\mathrm{(ours)}}-\sigma{\mathrm{(SDSS)}}$]$/
\sigma{\mathrm{(SDSS)}}$.
\label{berncomp}}
\vskip 5mm
%%%%%%%%%%%%%%%%%%%%%%%%%%%%%%%%%%%%%%%%%%%%%%%%%%%%%%%%%%%%%%%%%%%%%
\noindent
better reproduce the true stellar population mix (e.g.,~Rix \etal\
1995; Kobulnicky \& Gebhardt 2000), but such a treatment is beyond the
scope of this work.  Instead, we focus on 
well-defined spectral regions that are dominated by old stellar
populations in an effort to minimize template-mismatch effects.
Empirically, as found by, for instance, Barth \etal\ (2003, 2005),
early K giants consistently provide the closest match to both the
\mgb\ and \cat\ regions of many AGN samples.  Therefore, we believe we
are justified in this choice of stellar templates (we will examine
potential biases in more detail in \S 3.2.2).  One unfortunate
characteristic of these templates is their rather moderate S/N.  While
higher than the AGN spectra, they have a mean S/N of only 30, 75, and
65 in the regions surrounding \chk, \mgb, and \cat, respectively,
resulting in additional uncertainties from the velocity standards
themselves.  We have therefore adopted a rather unorthodox procedure.
Rather than performing an individual fit with each template star and
then representing our best fit as their average (e.g., Barth et
al. 2003), we create a master template by averaging all 32 stars
(hereafter the ``average template''), and use that as our velocity
standard.  To make the average template, we first align the spectra in
velocity (shifts of $\sim 0.1$ pixel or 7 \kms\ are typical) and then
compute a direct average.  We find from simulations that our error
bars are decreased using this average spectrum, while no systematic
uncertainties are introduced.  Likewise, our effective resolution is
not affected (see \S 3.2.2).

\subsection{Tests of the Direct-fitting Code}

As this is the first presentation of our velocity dispersion code, we devote 
extra care to examine the reliability and limitations of the method.  To begin
with, using comparisons with \sigmastar\ from the literature, we
demonstrate that we can reproduce established measurements.  We then
perform a suite of simulations, in which we measure the velocity
dispersion of an artificially velocity-broadened input model spectrum with 
noise and A star contamination added, in order to quantify the effect of these
variables on the accuracy and precision of our measurements.

\subsubsection{External Comparisons}

As an initial verification that our code functions reliably, we 
demonstrate that we can reproduce published stellar velocity
dispersions.  Heckman \etal\ (2004) present velocity dispersion
measurements for SDSS Type 2 AGNs, and the measurements for the 33,589
AGNs from the Second Data Release are made publicly available by
Brinchmann \etal\ (2004).  The velocity dispersions were measured from
the SDSS data using a direct-fitting algorithm to be described in
detail by D.~J.~Schlegel \etal\ (in preparation).  In short, their
template stars consist of the first few eigenspectra from a principal
component analysis of the echelle stellar spectra in the Elodie
database (Moultaka \etal\ 2004).  A Gaussian broadening function is
assumed, and the fitting is performed over the region from 4100 to
6800 \AA.  These form an excellent comparison sample for our work
because they are based on SDSS data using a very comparable method.
Note that because these are Type 2 AGNs, unlike our Type 1 sources,
their continua are dominated by starlight, although there may be some
nonstellar continuum as well (e.g.,~Zakamska \etal\ 2003).  As long as
we mask the emission features, the stellar continuum can be fitted in
a straightforward manner.  In order to select reasonably sized
subsamples of only the closest galaxies at each value of \sigmastar,
we applied different redshift constraints in different \sigmastar\
bins.  For our chosen bins in velocity dispersion we employed the
following (rather ad hoc) redshift cuts: 70--125 \kms, $z \leq 0.025$;
125--150 \kms, $z \leq 0.03$; 150--170 \kms, $z \leq 0.035$; 170--225
\kms, $z \leq 0.04$; 225--280 \kms, $z \leq 0.05$; and 280--400 \kms,
$z \leq 0.1$. After removing all galaxies with a S/N $< 20$, we are
left with a total of 504 objects.  We fit a slightly different
spectral region, from 4000 to 6200 \AA, but we find excellent
agreement with the SDSS measurements, as shown in Figure 2.  The mean
in the residuals, when expressed as a fractional difference, is 
$\langle$\ldelsig$\rangle$=$\langle \mathrm{[} \sigma{\mathrm{(ours)}} -
\sigma{\mathrm{(SDSS)}} \mathrm{]}/\sigma{\mathrm{(SDSS)}} \rangle = -0.016 \pm
0.13$.  Although the absolute scatter increases at the highest
velocity dispersions, in relative terms the scatter is dominated by
the lower velocities, particularly close to the resolution limit.  
Dividing the sample, we find $\langle$\ldelsig$\rangle=-0.018 \pm 0.14$ for
\sigmastar\ $\leq 200$ \kms\ versus
$\langle$\ldelsig$\rangle=-0.013 \pm 0.087$ at larger \sigmastar.

We also use our comparison with the SDSS sample to explore potential
uncertainties introduced by the polynomial order used in the fit.  Our
standard fit employs a third-order polynomial.  If we lower the order
to 2 we find a systematic offset of roughly $-5 \%$ from the SDSS
values.  Even worse, when we use a fourth-order polynomial we find
large systematic differences.  At low \sigmastar\ ($\leq 200$~\kms) we
find a $\sim 25 \%$ positive offset, while there is reasonable
agreement for \sigmastar\ $\geq 350$ \kms, very similar to 
the trend reported by Barth \etal\ (2002a).  Naively, we had expected
the large spectral range to minimize the importance of the polynomial
order.  However, as noted by Barth \etal, since our spectra are
flux-calibrated, high-order polynomial terms may begin to fit

%%%%%%%%%%%%%%%%%%%%%%%%%%%%%%%%%%%%%%%%%%%%%%%%%%%%%%%%%%%%%%%%%%%%
%%BoundingBox: 18 144 592 680
\psfig{file=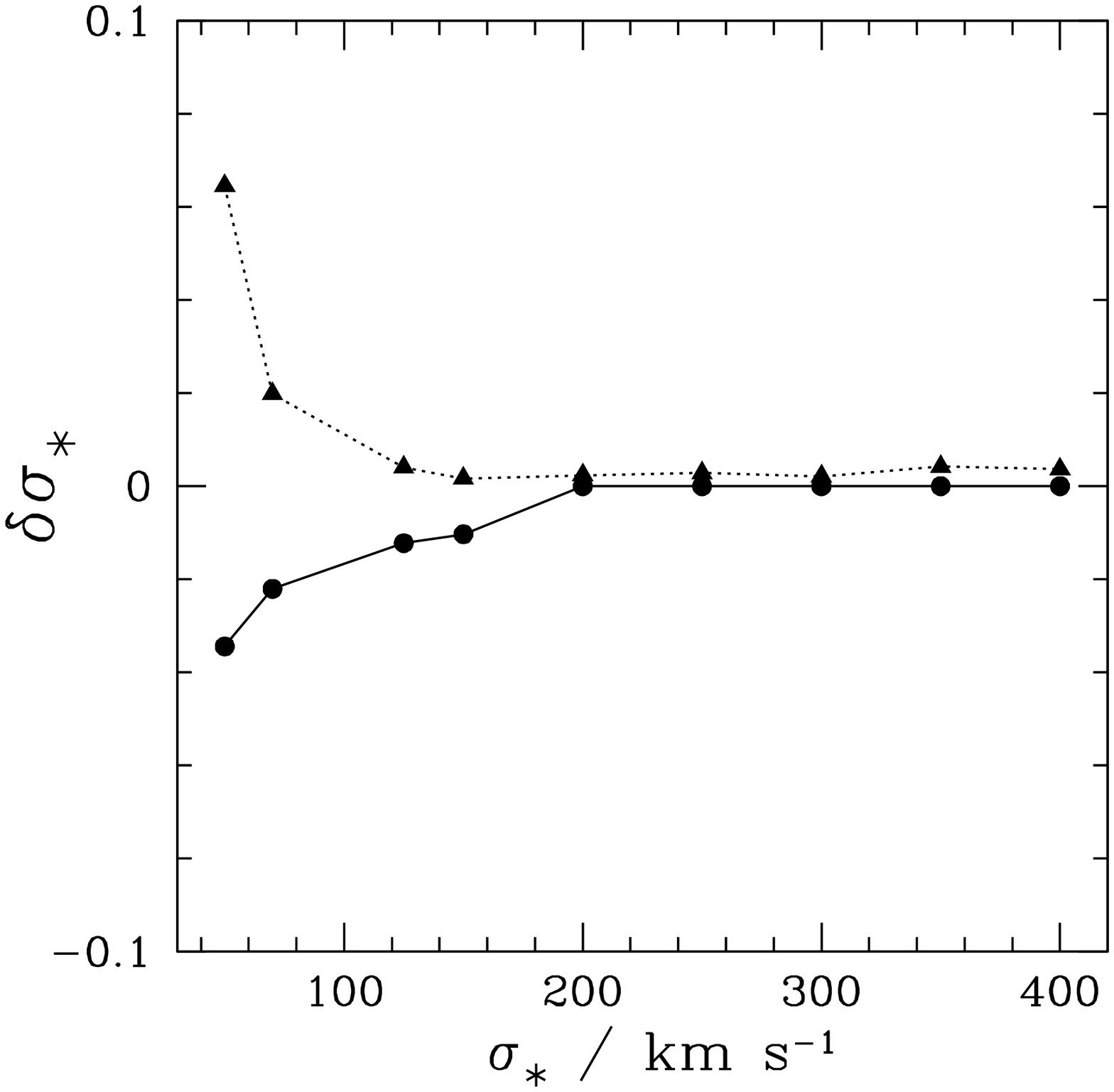,width=0.47\textwidth,keepaspectratio=true,angle=0}
\vskip -4mm
\figcaption[]{Input-output test using an average of 10 template stars
as a model galaxy.  \ldelsig $\equiv$
[$\sigma{\mathrm{(out)}}-\sigma{\mathrm{(in)}}$]$/\sigma{\mathrm{(in)}}$.
The triangles are derived from the average of the fit from each
individual template star, while the circles come from fitting with the
average template.  The results are similar, but the average template
behaves better at the lowest velocity dispersions.  We see that the
effect of resolution becomes noticeable for \sigmastar\ \lax\ 100 \kms.
\label{avgvall}}
\vskip 5mm
%%%%%%%%%%%%%%%%%%%%%%%%%%%%%%%%%%%%%%%%%%%%%%%%%%%%%%%%%%%%%%%%%%%%
\noindent
individual spectral features.  To minimize this systematic effect, in
the following we restrict our fits to polynomials of order 3.

As an additional check, because our code is newly developed
for this work, we have compared our results with the measurements of
two well-established direct-fitting codes.  M. Sarzi kindly 
provided an IDL adaptation of the Rix \& White (1992) code for
intercomparison.  For a subsample of 13 high-S/N, pure 
absorption-line (early-type) galaxies taken from the SDSS sample of
Bernardi \etal\ (2003) we found, for \ldelsig\ $\equiv$
[$\sigma{\mathrm{(ours)}}-\sigma{\mathrm{(test)}}$]$/\sigma{\mathrm{(test)}}$,
that $\langle$\ldelsig $\rangle = 0.01 \pm 0.03$.  Since the Rix \&
White code was not designed specifically to deal with AGN
contamination, we also compared our results with those obtained using
the code presented in Barth \etal\ (2002a).  A. J. Barth kindly 
used his code to fit six of the objects from our sample over an identical
fitting region. Again, our results agreed within the uncertainties,
with no systematic differences apparent, yielding $\langle$\ldelsig
$\rangle = 0.01 \pm 0.05$, where we attribute the small differences to
differences in the minimization routines.  These tests give us
additional confidence in the reliability of our code.

\subsubsection{Systematics}

%%%%%%%%%%%%%%%%%%%%%%%%%%%%%%%%%%%%%%%%%%%%%%%%%%%%%%%%%%%%%%%%%%%%
\begin{figure*}[t]
\vbox{ 
\vskip -0.6truein
\hbox{
\psfig{file=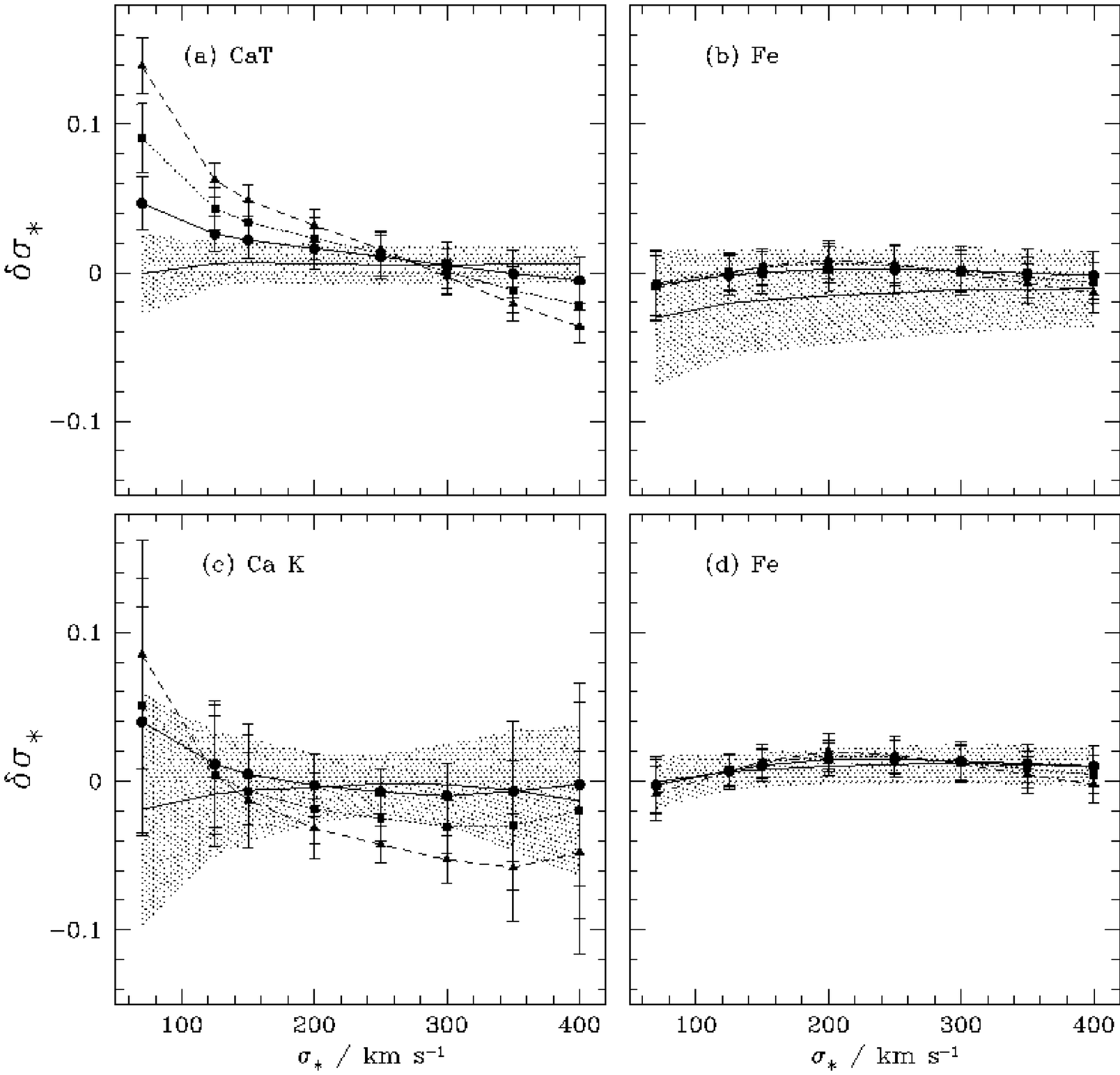,width=0.88\textwidth,
keepaspectratio=true,angle=0}
}} 
\vskip +15mm
\figcaption[]{ 
Uncertainty, expressed as a fractional difference \ldelsig $\equiv$
[$\sigma{\mathrm{(out)}}-\sigma{\mathrm{(in)}}$]$/\sigma{\mathrm{(in)}}$,
due to template mismatch for ({\it a}) CaT, ({\it b}) the Fe
region (5250--5820 \AA), and ({\it c}) Ca~K.  We create 100
different model galaxies from random collections of 10 template
stars, and depict one standard deviation above and below the mean
\ldelsig\ by the shaded region.  The mean is demarcated by the thin
solid line.  We also run simulations with 10\%
({\it circles}), 20\% ({\it squares}), and 30\% ({\it triangles}) A star 
contribution.  The points represent the mean \ldelsig, and the error bars are 
one standard deviation in \ldelsig.  Because of the strange behavior of Fe
shown in ({\it b}), we present alternate simulations in ({\it d}) excluding 
the six largest outliers.  
\label{templmismatch}}
%\vskip 5mm
\end{figure*}
%%%%%%%%%%%%%%%%%%%%%%%%%%%%%%%%%%%%%%%%%%%%%%%%%%%%%%%%%%%%%%%%%%%%

Many factors may contribute to uncertainties in our \sigmastar\
measurements, including template mismatch, resolution, and S/N (we discuss
emission from the AGN separately).  By generating model spectra, we can
examine how each of these factors in turn impacts our ability to
recover the input velocity width.  We build model galaxies as linear
combinations of template spectra, which we then broaden to a range of
widths between $\sigma$ = 50 and 400 \kms, assuming a Gaussian line-of-sight
velocity distribution.  We then vary parameters of interest, such as S/N, and 
evaluate our ability to recover \sigmastar, using as the figure of merit the
fractional error \ldelsig\ $\equiv$
[$\sigma{\mathrm{(out)}}-\sigma{\mathrm{(in)}}$]/$\sigma{\mathrm{(in)}}$.
For later use, we focus on (1) the region around \chk\ (3910--4060 \AA) with 
Ca~H (3955--3985 \AA) masked, (2) the region redward of \mgb\ (5250--5820 \AA; 
hereafter called the ``Fe region'') with [\ion{Fe}{7}]~$\lambda 5721$ masked, 
and (3) the region around the \cat\ (8470--8700 \AA). Each fit is done locally, 
without attempting to match the polynomial or AGN power-law between regions.

%%%%%%%%%%%%%%%%%%%%%%%%%%%%%%%%%%%%%%%%%%%%%%%%%%%%%%%%%%%%%%%%%%%%
\begin{figure*}[t]
\vbox{ 
\vskip -0.0truein
\hbox{
\psfig{file=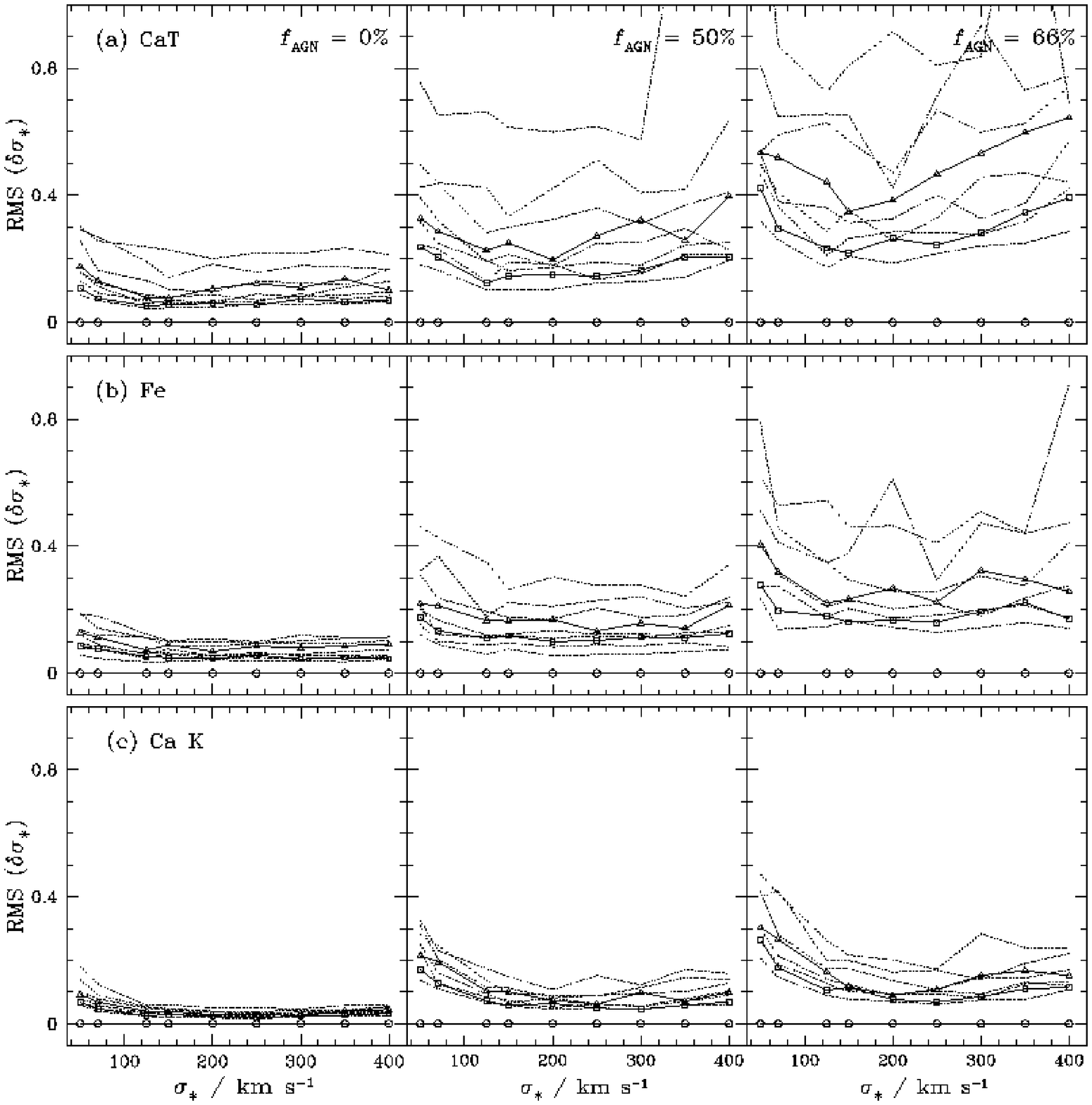,width=18.cm,angle=90} 
}} 
\figcaption[] {
Input-output simulations of velocity dispersion measurements from
({\it a}) the \cat\ region, ({\it b}) Fe (5250--5820 \AA), and ({\it
c}) Ca~K of galaxy models constructed as the average of 10 template
stars.  \ldelsig\ $\equiv$
[$\sigma{\mathrm{(out)}}-\sigma{\mathrm{(in)}}$]$/\sigma{\mathrm{(in)}}$.
Simulations are shown for S/N of 15 to 55, in steps of 5.
The highest S/N (i.e., no additional noise) is $\sim$ 205, 240, and 95
for \cat, Fe, and Ca~K, respectively (shown in {\it open circles}).
We highlight in solid lines the cases for S/N = 30 ({\it triangles})
and S/N = 50 ({\it squares}), since the majority of our data are
spanned by these limits.  The {\it middle} and {\it right}\ columns correspond 
to $f_{\rm AGN}$ = 50\% and 66\%, respectively.
\label{fakecat}}
%\vskip +5mm 
\end{figure*}
%%%%%%%%%%%%%%%%%%%%%%%%%%%%%%%%%%%%%%%%%%%%%%%%%%%%%%%%%%%%%%%%%%%%

To begin with the simplest possible test, we fit a model galaxy,
composed of 10 template stars with only artificial broadening, in the
\cat\ region.  We are using a single average template in order to
increase the effective S/N of our template, but for comparison we also
compute \sigmastar\ as the average of the values obtained from each of
the 32 individual template stars. The resulting \ldelsig\ values for
both approaches are shown in Figure 3.  No apparent offset is
introduced by the use of an average template, while the fractional
error close to the resolution limit is diminished.  In general, while
additional uncertainty is introduced below \sigmastar\ $\approx 100$
\kms\ due to resolution, it is only on the order of $\sim 2 \%$.  We follow 
previous work (e.g.,~Bernardi \etal\ 2003; Heckman \etal\ 2004) and limit our 
attention only to velocity dispersions larger than the nominal 
instrumental resolution of SDSS, which we take to be 
$\sigma_{\rm inst} = 70$ \kms.

Even for predominantly old stellar populations (G and K giant stars),
template mismatch will introduce feature-dependent uncertainties.  To
investigate the magnitude of these uncertainties for each fitting
region, we generate a family of 100 model galaxies.  As above, the
model consists of 10 templates, which are selected randomly for each
run, and then coadded and artificially broadened.  The shaded region
in Figure 4 represents one standard deviation around \ldelsig\ for the
ensemble of 100 trials.  Note that the \cat\ region is well-behaved (Fig. 
4{\it a}), while the Fe region is, on average, $\sim 2\%$ low (Fig. 4{\it b}).
However, when we remove the six largest outlying templates from the
parent set of 32 and repeat the simulations, the negative offset is
removed (Fig. 4{\it d}).  In the \chk\ region \ldelsig\ has a mean
value of zero, but by far the largest dispersion.  

While this experiment gives us a feel for template mismatch among old
stellar populations, additional uncertainty is introduced by young
stellar populations.  Particularly, it has been shown that the host
galaxies of luminous AGNs do not contain a uniformly old population,
but rather are more likely to contain a post-starburst population than
quiescent galaxies of similar total stellar mass (e.g.,~Terlevich
\etal\ 1990; Kauffmann \etal\ 2003b; Nelson \etal\ 2004).  Due to
rotational and pressure broadening, a significant A star population
can significantly broaden the hydrogen lines, and potentially bias the
measurements in the regions around \cat\ and \chk.  Therefore, we run
an additional set of simulations in which we explicitly include an A
star contribution, using an A star spectrum observed by SDSS,
Gaussian-broadened to the same width.  In Figure 1 (bottom panel) we
show that the A star spectrum has significant features in both the
\chk\ and \cat\ regions, while it is nearly featureless in the Fe
region.  In the case of \cat, the features are a blend of Paschen and
\cat\ lines, while the Ca K line is apparent to the left of a blend of
Ca H and H$\epsilon$ in the \chk\ region.  The
\ldelsig\ values for models including A star contributions of 10\%,
20\%, and 30\% are shown in Figure 4.  As expected, both the \chk\ and
\cat\ regions are more sensitive to the presence of A stars, although
the impact is relatively small ($\leq 10\%$).  Note, however, that the
induced spread is largest in the case of \chk.

The situation is more puzzling in the case of the Fe region, where the
uncertainties appear to {\it decrease} with the addition of an A star
component.  Apparently an A star contribution leads to a better fit of
the outlying templates, which were removed to generate Figure 4{\it
d}.  Various scenarios may account for this behavior.  If, for
instance, the outliers have a continuum shape opposite to that of an A
star (i.e., very red), then the addition of an A star would cancel
them out.  We investigated this possibility by running
simulations including the A star but excluding the outlying templates,
shown as points in Figure 4{\it d}, but the induced errors do not
increase, so this is probably not the solution.  Another possibility
is that the outlying templates have stellar absorption features with
larger EWs, so that the additional dilution from the A star continuum
improves their fit with the rest of the templates.  We measured
the EW in \mgb\ (using the Lick index definition; e.g.,~Trager \etal\
1998) for each individual template, and the mean EW for the outliers
is 5.2 \AA, compared to $\langle$EW$\rangle$ = 3.3 \AA\ for the rest,
indicating this is the most likely explanation.  Since our actual
galaxies will probably contain some A star contribution, we have
decided to retain all 32 templates in our average template, despite
the apparent skewness of Figure 4{\it b}.  We verified that our
actual fits are not strongly affected if we use an average template
excluding the six outliers.  Between the nominal results and those
derived from an average template excluding the six outliers, we find
an average relative difference of $\langle$\ldelsig$\rangle = -0.06
\pm 0.1$.

We now turn to the impact of varying S/N on the reliability of our
measurements.  For these simulations, we begin with a well-behaved
model galaxy from the simulation above.  We degrade the S/N by adding
mean-zero Gaussian random deviates, whose uncertainty at each pixel is
given by the SDSS pipeline error array, and whose amplitude is varied
to decrease the effective S/N from 60 to 15, in steps of 5.  Note that
the model galaxy has a S/N of $\sim$ 205, 240, and 95 for the \cat,
Fe, and \chk\ regions, respectively, which forms the upper limit on
our noise simulations.  In practice, the majority of our objects are
found between a S/N of 30 and 50 (although the distribution is
somewhat lower in the Ca H+K region).  For each S/N, we generate 50
model spectra, in order to average over the variations caused, for
instance, when a noise spike falls on a crucial absorption feature.
We do not find a systematic bias in \ldelsig\ from variations in S/N
or dilution and so for simplicity we report the standard deviation in
\ldelsig over the 50 trials, denoted RMS(\ldelsig).
The results are shown in the
left-hand panels of Figure 5, for each spectral region of interest.
From these simulations it is clear that even at the modest S/N of 20,
which is common for SDSS spectra, it is possible to achieve
\sigmastar\ measurements with $\sim 20\%$ accuracy.

\vskip 1.0cm
\section{Measuring \sigmastar\ in AGNs}

Inspired recently by the discovery of the \msigma\ relation, many
groups have measured \sigmastar\ in Type 1 AGNs (e.g.,
Jim{\'e}nez-Benito \etal\ 2000; Ferrarese \etal\ 2001; Barth
\etal\ 2002b, 2003, 2005; Nelson \etal\ 2004; Onken \etal\ 2004; Treu
\etal\ 2004; Woo \etal\ 2004; Botte \etal\ 2005; Garcia-Rissmann
\etal\ 2005).  As noted by many of these authors, measuring
\sigmastar\ in the presence of an AGN is particularly challenging.
The AGN power-law continuum dilutes the stellar features, effectively
lowering the S/N, while emission lines from the AGN can add
(sometimes subtle) biases into the fits.  In order to mitigate these
problems, the above studies fit specific spectral regions with minimal
AGN contamination but high-EW stellar absorption features, rather than
attempting to find global solutions for large regions of the spectrum.
We adopt the same approach.  In particular, the utility of the \cat\
region has long been recognized (Pritchet 1978; Dressler 1984).  The
\cat\ lines are intrinsically narrow compared to galactic velocity
dispersions, with intrinsic widths of $\sigma \approx 20-25$ \kms\ for
K giant stars, ranging up to $\sim 30$~\kms\ for supergiant stars (Filippenko
\& Ho 2003; Martini \& Ho 2004).  Furthermore, the AGN continuum is
minimized in their vicinity, since AGN continua are typically blue.
Although higher-order Paschen lines (P13~$\lambda 8503$, P15~$\lambda 8545$, 
and P16~$\lambda 8665$) overlap with the \cat\ lines, these
features are rarely seen in emission (an exception is
NGC 4395, which, as noted by Filippenko \& Ho, has anomalously high-EW
emission features).  Emission from \cat\ itself can fill in the
absorption features as well, but these objects were eliminated by
selection.  Finally, there may also be \ion{O}{1}~$\lambda 8446$ or
[\feii]~$\lambda 8616$ emission from the AGN.  In principle we can
mask either of these features in a straightforward manner within our
fit, but again, these lines are weak by selection in our sample.  We
measured meaningful \sigmastar\ measurements for all but one of our
objects, although an additional four are below our nominal resolution
limit of $\sim 70$ \kms\ (Table 1).  At the S/N of the SDSS data, our
ability to recover accurate \sigmastar\ measurements below the
resolution limit is significantly decreased, and so we 
conservatively do not include
these four measurements in our analysis.

Of all high-EW features in the optical bandpass, the \cat\ lines are
simultaneously the least affected by template mismatch and AGN
contamination, making them our ``gold standard.''  Unfortunately,
because they occur in the red region of the spectrum, they are
inaccessible to optical bandpasses above $z \approx 0.05$.  For this
reason, we must examine the utility of other spectral regions with
reasonably strong absorption features.  We will focus on two: the
region surrounding the \mgb\ triplet (5040--5820 \AA) and the region
around the Ca H+K lines (3900--4060 \AA).  We measure \sigmastar\
using these alternate spectral regions, and compare the results
directly with the \cat\ values.  In order to evaluate whether a given test
region matches \cat, we use the same figure of merit as
defined above, \ldelsig\ $\equiv$
[$\sigma{\mathrm{(test)}}-\sigma{\mathrm{(\cat)}}$]/$\sigma{\mathrm{(\cat)}}$,
using in all cases the 35 measurements with \sigmastar(\cat) $\geq
70$~\kms.

\subsection{Power-law Continuum Contamination}

Relatively speaking, it is straightforward to model the contaminating
effect of a pure power-law component on the final \sigmastar\
measurement.  We perform simulations identical to those described in
\S 3.2.2, starting with a model composed of 10 template stars,
adjusting the S/N from 15 to 60, in steps of 5, and measuring
\ldelsig, except that now we also include a power-law component to
represent the AGN continuum.  We use a single 
power law, $f_\nu
\propto \nu^{-1}$, comprising 0, 33, 50, 60, 66, 70, 75, 80\% of the
total local continuum, as measured in the 20 \AA\ at each end of the
band.  Assuming our nominal power-law slope of $-1$, an AGN fraction of
$50\%$ ($66\%$) in the Fe region corresponds to 

%%%%%%%%%%%%%%%%%%%%%%%%%%%%%%%%%%%%%%%%%%%%%%%%%%%%%%%%%%%%%%%%%%%%
%%BoundingBox: 102 46 670 724
%\vskip 10mm
%\hskip 5mm
\psfig{file=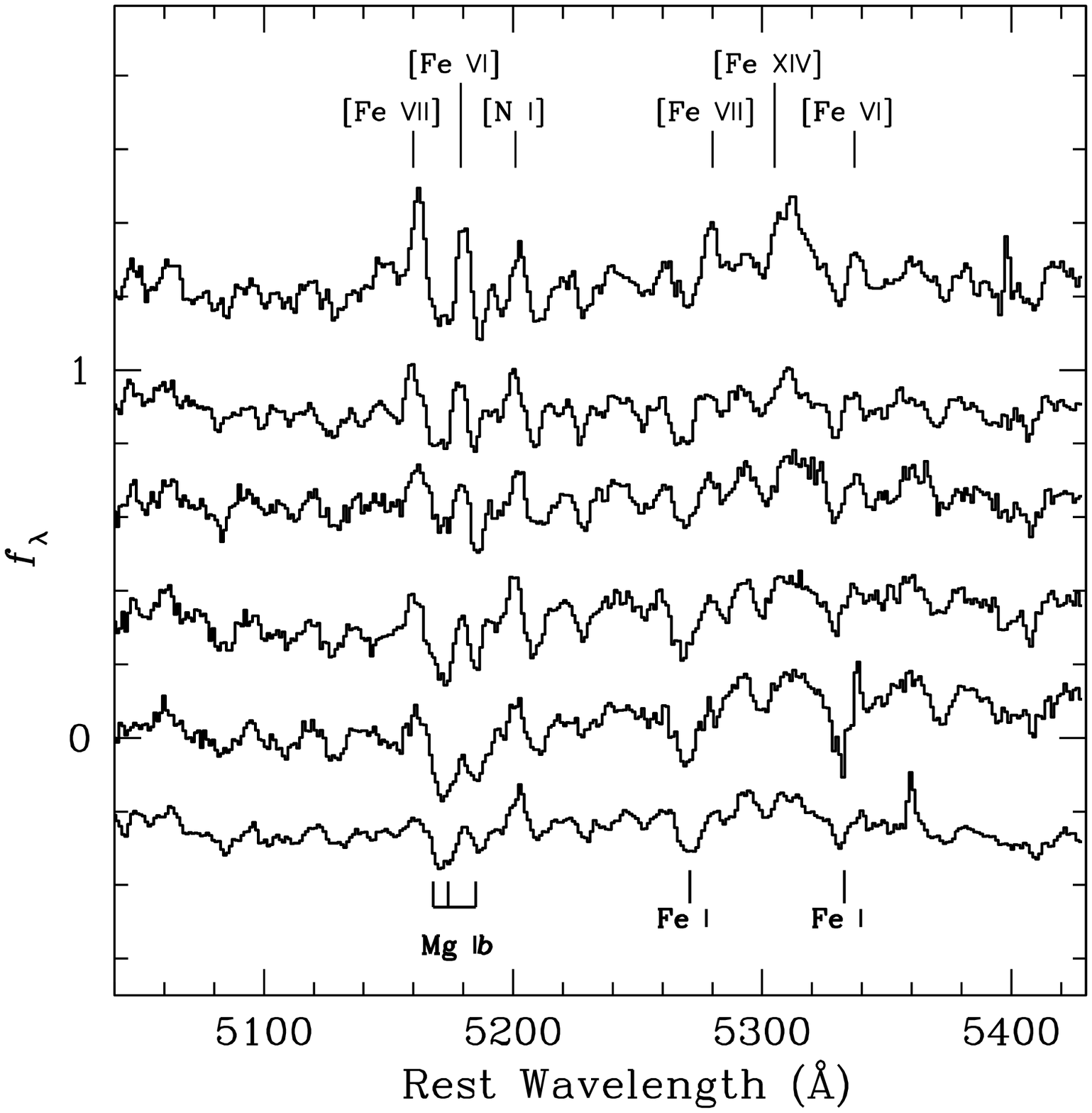,width=9.5cm,angle=0}
\vskip -18mm
\figcaption[]{
Examples of the Mg~{\tiny I}{\it b} spectral region for objects with
significant Fe~{\tiny II} contamination, as well as, in some cases,
contamination from
[Fe~{\tiny VII}]$~\lambda 5158$, [Fe~{\tiny VI}]~$\lambda 5176$,
[N~{\tiny I}]~$\lambda \lambda$5197, 5177,
[Fe~{\tiny VII}]$~\lambda 5278$, [Fe~{\tiny XIV}]$~\lambda 5303$, and
[Fe~{\tiny VI}]$~\lambda 5335$.  Plotted in order of decreasing
[Fe~{\tiny VII}] strength, from top to bottom, are SDSS J162012.75+400906.1,
J120114.35$-$034041.0, J083949.65+484701.4, J112536.16+542257.1, 
J150745.00+512710.2, and J101912.56+635802.6.
\label{fespec}}
\vskip 5mm
%%%%%%%%%%%%%%%%%%%%%%%%%%%%%%%%%%%%%%%%%%%%%%%%%%%%%%%%%%%%%%%%%%%
\noindent
$48\%$ ($ 65\%$) AGN
fraction in the \cat\ region and $73\%$ ($84\%$) AGN fraction in the Ca~H+K 
region.  For general comparison with the literature, we also compute the
EW in \mgb, \cat, and \chk\ from the average template star for AGN
fractions of $0\%, 50\%$, and $66\%$ in the Fe region.  We use the
Lick index definition of the \mgb\ index (Trager \etal\ 1998), the
Terlevich \etal\ (1989) definition of the \cat\ $\lambda 8542$ \AA\
line, and the Brodie \& Hanes (1996) definition of the HK index.  For
no dilution we find EW = 4, 3, and 22 \AA\ for \mgb, \cat, and \chk,
respectively.  These decrease to EW = 2, 1, and 6 \AA\ for $f_{\rm
AGN}$ = 50\%, and to EW = 1, 0.9, and 4 \AA\ for $f_{\rm AGN}$ = 66\%.
We show representative cases with 50\% and 66\% contamination in the
middle and right panels of Figure 5.  Although we have run a larger
suite of simulations, these examples bracket the majority of the measured
values for our sample.  We assign an uncertainty to each of our
galaxies by interpolating between the grid points from the simulations
to its measured S/N and dilution.  These simulations show that if
template mismatch and AGN emission features were not present, we would
achieve the best measurements using the line with the highest EW
(namely \chk\ in this case).  While not directly relevant to this
particular sample, it is of general interest to determine the limiting
AGN dilution for each spectral feature.  As a general indicator, we
set the maximum AGN dilution at the point where \ldelsig\ = 1 at a S/N
= 20.  This occurs at an AGN fraction of 71\% for \cat, 85\% for the
Fe region, and 90\% for \chk.  Although the \chk\ lines endure the
highest AGN contamination, they are very susceptible to systematic
uncertainties induced by template mismatch, and so we prefer the \cat\
lines despite their lower EWs.

As a sanity check, we compare to the BL Lac objects studied by Barth
\etal\ (2003), whose dominant sources of uncertainty are S/N and
dilution from a pure power-law continuum.  For their sample, with
dilutions of 50\%--80\% and S/N between 50 and 300, they find
$\sim$2\%--11\% errors using either the \cat\ or \mgb\ spectral
regions.  Reassuringly, to the extent that they overlap, our
simulations are consistent with their results.

\subsection{Emission-line Contamination}

%%%%%%%%%%%%%%%%%%%%%%%%%%%%%%%%%%%%%%%%%%%%%%%%%%%%%%%%%%%%%%%%%%%%
\begin{figure*}[t]
\vbox{
%\vskip -0.08truein
\hbox{
\psfig{file=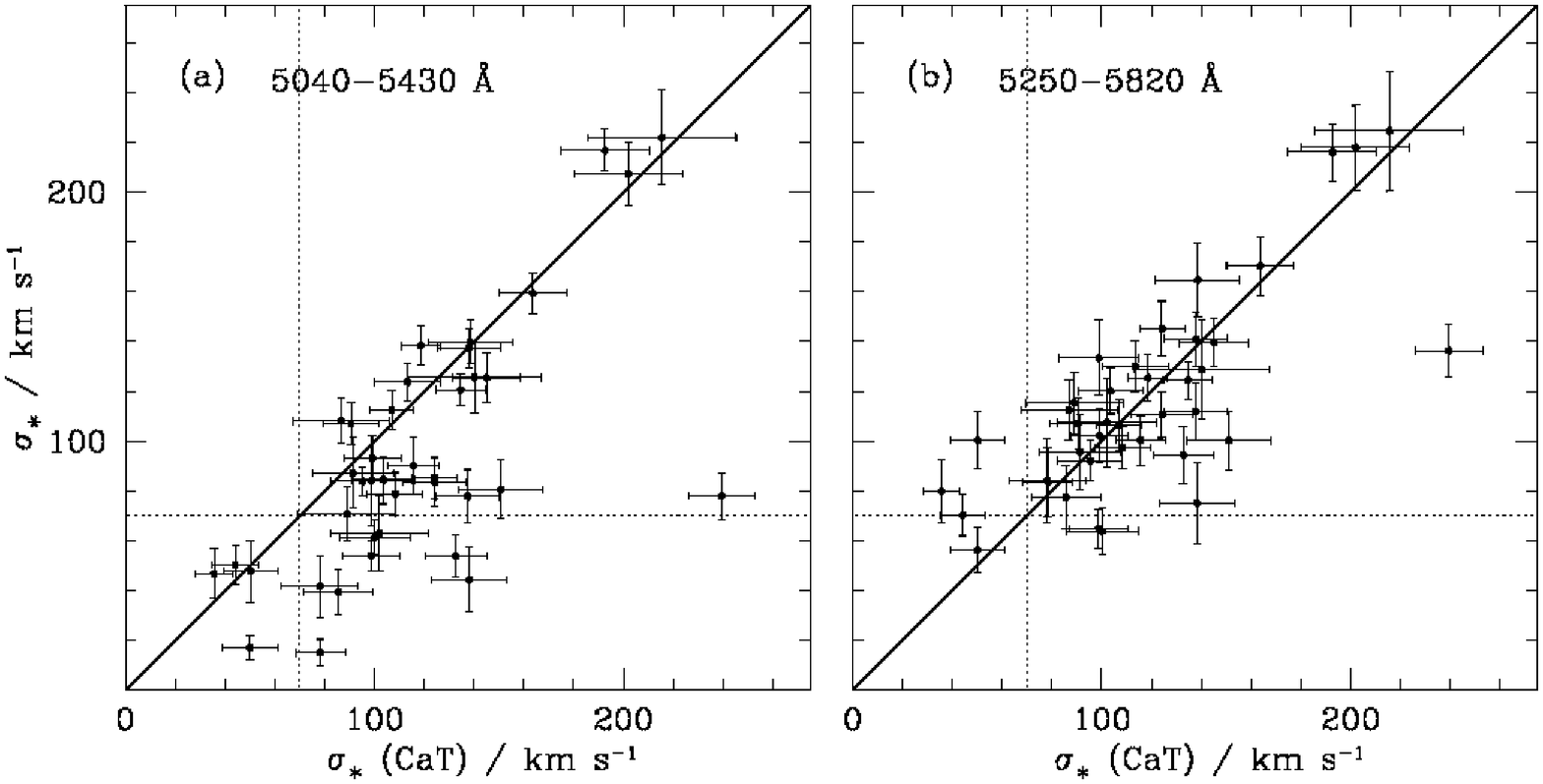,width=0.47\textwidth,angle=-90}
}}
\vskip +5mm
\figcaption[]{
A comparison between $\sigmastar$ from the \cat\ region and ({\it a}) the
Mg {\tiny I}{\it b} region (5040--5430 \AA) and ({\it b}) the Fe region
(5250--5820 \AA).  Note the large systematic offset at low \sigmastar\ in the
Mg {\tiny I}{\it b} region.  The solid line represents equality,
while the dotted lines mark the nominal resolution limit of the SDSS.
\label{sigcomp}}
\end{figure*}
%%%%%%%%%%%%%%%%%%%%%%%%%%%%%%%%%%%%%%%%%%%%%%%%%%%%%%%%%%%%%%%%%%%%

As we have seen, contamination from a power-law continuum only
effectively lowers the S/N of the measurement.  Far more pernicious
are the subtle biases that may be introduced by AGN emission features
and template mismatch.  Between the \mgb\ lines themselves and the
prominent \ion{Fe}{1} blends at 5270~\AA\ and 5335~\AA, this region
contains sufficiently high-EW features for measuring velocity
dispersion, but they are susceptible to template mismatch due to
metallicity effects.  Because of the well-known correlation between
[Mg/Fe] and $\sigmastar$ (e.g.,~O'Connell 1976; Worthey \etal\ 1992;
Trager \etal\ 1998; J{\o}rgensen 1999; Kuntschner \etal\ 2001), it
becomes increasingly difficult to simultaneously fit both the \mgb\
and Fe features using template spectra with solar metallicities and
abundance ratios.  The errors incurred are small, but increase with
increasing $\sigmastar$, tending to artificially increase the inferred
velocity dispersion (Barth et al. 2002a).  

Perhaps more relevant to
this study, there is significant AGN line contamination over this
wavelength range.  There are a number of narrow emission lines,
including [\ion{Fe}{7}]$~\lambda 5158$, [\ion{Fe}{6}]~$\lambda 5176$,
[\ion{N}{1}]~$\lambda \lambda$5197, 5200, [\ion{Fe}{7}]$~\lambda
5278$, [\ion{Fe}{14}]$~\lambda 5303$, and [\ion{Fe}{6}]$~\lambda 5335$
(see Fig. 6).  The first three lines are especially troublesome,
occurring as they do amidst the \mgb\ features, and they are present
in $\sim 50\%$ of our sample.  In addition to narrow features, there
is a broad \feii\ ``pseudo-continuum,'' extending from $\sim$5050 to
5520 \AA\ (e.g.,~Francis \etal\ 1991).  When the \feii\ emission is
broad enough to be smooth, it simply lowers the effective S/N of the
spectrum without introducing significant systematic errors
(e.g.,~Nelson \& Whittle 1995).  However, when the broad-line velocity
is low enough that the individual \feii\ components are relatively
resolved, then the resulting \sigmastar\ measurements may be biased.
Some examples of galaxies with significant broad \feii\ contamination
as well as strong coronal emission lines are shown in Figure 6.

Following Barth \etal\ (2002a), we use the region extending 5040 to
5430 \AA, and we mask the [\ion{N}{1}] feature from 5190--5210 \AA.
In Figure 7{\it a}, we compare the velocity dispersions obtained from
this fit with those from \cat.  As is immediately clear,
$\sigma$(\mgb) is significantly biased.  In particular, there is a
plume of objects with $\sigma{\mathrm{(CaT)}}$ values between 75 and
150 \kms\ and $\sigma{\mathrm{(Mg {\tiny I}{\it b})}} \approx 80$ \kms.  We 
find \ldelsig\ = $\langle \mathrm{[} \sigma{\mathrm{(Mg {\tiny I}{\it b})}}
-\sigma{\mathrm{(CaT)}} \mathrm{]} /\sigma{\mathrm{(CaT)}} \rangle$ = $-$0.23
$\pm$ 0.32.  A combination of
broad, permitted \feii\ emission and high-ionization forbidden
emission must be responsible for this dramatic offset, since
mismatches in [Mg/Fe] tend to go in the opposite direction.  We
use simulations to gauge the systematic effects due to \feii\
contamination, and we experiment with different spectral regions to
minimize the effects of coronal emission.

We examine the role of broad \feii\ contamination by running
simulations, similar to those presented in \S 3.2.2.
In the usual way, we create a model
galaxy using 10 of our template stars and Gaussian-broaden it
artificially to widths between 50 and 400 \kms.  As a model of the
\feii\ contamination we utilize the \feii\ spectrum of \zw\ (Boroson
\& Green 1992), kindly provided by T. A. Boroson.  \zw\ is a narrow-line
Seyfert 1 galaxy, which for our purposes ensures that the \feii\
emission is both strong and narrower than the \feii\ emission in our
objects.  The challenge is
simulating a representative range of EWs and velocity dispersions for
the \feii\ template that accurately reflect what is found in real
AGNs.  In the sources of interest the \mgb\ features are strong (and
thus measurable), making it difficult to accurately determine the
\feii\ EW, at least in a model-independent way.  We have chosen to
vary the \feii\ amplitude between an EW of 18 and 54 \AA, comparable
to what is seen in relatively weak \feii\ emitters.  For instance,
Boroson \& Green (1992) find a range of \feii\ EWs between 0 and 114
\AA\ for their sample of nearby quasars.  As demonstrated in Figure 8,
this level of \feii\ contamination is hard to discern based on the
appearance of the absorption features.  However, despite the low
amplitude, this quantity of \feii\ contamination can seriously bias
the fits.  Far more important than the amplitude of contamination is
the width of the \feii\ lines.  When they are narrow
enough to have significant structure, they most effectively confuse the fitting
procedure.  Empirically, the width of the \feii\ emission is
found to match that of the broad component of \hbeta\ (e.g.,~Boroson
\& Green 1992).  Therefore, using the \msigma\ relation, for any given
\sigmastar\ the range of possible \feii\ widths is set by the
luminosity of the AGN through the line width-luminosity relation
(\mbh\ $\propto$ $L^{0.64}$\fwhb$^2$; Kaspi \etal\ 2005; Greene \& Ho
2005b).  Since, for a given \mbh\ the maximum luminosity is set by the
Eddington limit, it is convenient to parameterize the AGN luminosity
using the Eddington ratio, \lledd, where $L_{\mathrm{Edd}} \equiv 1.26
\times 10^{38}$~(\mbh/\msun) \lum.  In practice, the bolometric
luminosity is difficult to obtain, and so we assume a canonical
bolometric correction of 0.1 between the optical and bolometric
luminosities (e.g.,~Elvis \etal\ 1994).  For a fixed \sigmastar\ (or
equivalently \mbh), as the luminosity of the system increases the line
width must decrease, such that the narrowest \feii\ lines are found at
the highest \lledd.  Hence, the worst contamination occurs at the
Eddington limit, as shown in Figure 9, where we consider \lledd\ =
0.01, 0.1, and 1.  Also, for any given value of \lledd, the bias is
worst near the resolution limit.\footnote{This explains why, for
instance, Barth \etal\ (2005) see no bias between their Mg{\tiny
I}{\it b} and \cat\ measurements, since they are measuring velocity
dispersions 3--5 times the effective spectral resolution.}  We do not
show the simulations with \feii\ EW of 18 or 54 \AA, but the general
trends are the same, with a maximum bias in the \mgb\ region of $\sim
10\%$ for an EW of 18 \AA\ and $\sim 30\%$ for an EW of 54 \AA. In
addition to the 5040--5430 \AA\ region, we also consider the region
5250--5820 \AA, which eliminates the \mgb\ features themselves but
includes the \ion{Fe}{1} absorption features.  As we will show below,
this is a better spectral region to use in cases with strong
[\ion{N}{1}] and coronal emission from [\ion{Fe}{7}].  We find that
this spectral region is somewhat better behaved at low \sigmastar\ and
low Eddington ratio, but
worse at large \sigmastar\ and at high \lledd\ (Fig. 9{\it b}).
Finally, we consider the effects of \feii\ contamination on the \chk\
lines.  The \feii\ features near $\sim 3950$ \AA\ are both weaker and
broader (see bottom of Fig. 1), which means that the maximum incurred
bias is $\leq 20\%$ (Fig. 9{\it c}; $\sim 10\%$ for EW of 18 \AA\ and
$\sim 30\%$ for an EW of 54 \AA), except near the resolution limit.
We note that while we have run a suite of simulations including 
many combinations of \lledd\ and EW, in nature the \feii\ strength is
found to increase with \lledd\ (e.g.,~Boroson 2002), further
contaminating sources close to their Eddington limit.

Given the rather large bias that \feii\ emission can apparently cause,
especially at high Eddington ratios, we seek a general method to
salvage the \mgb\ region by including an \feii\ model directly in our
direct-fitting code.  Equation 1, describing our fitting procedure,
includes a term $C(x)$, which is a model of the AGN continuum.
Previously we have modeled the AGN component as a pure power law, but
now we also include the \zw\ 

%%%%%%%%%%%%%%%%%%%%%%%%%%%%%%%%%%%%%%%%%%%%%%%%%%%%%%%%%%%%%%%%%%%%
%%BoundingBox: 88 144 592 718
\psfig{file=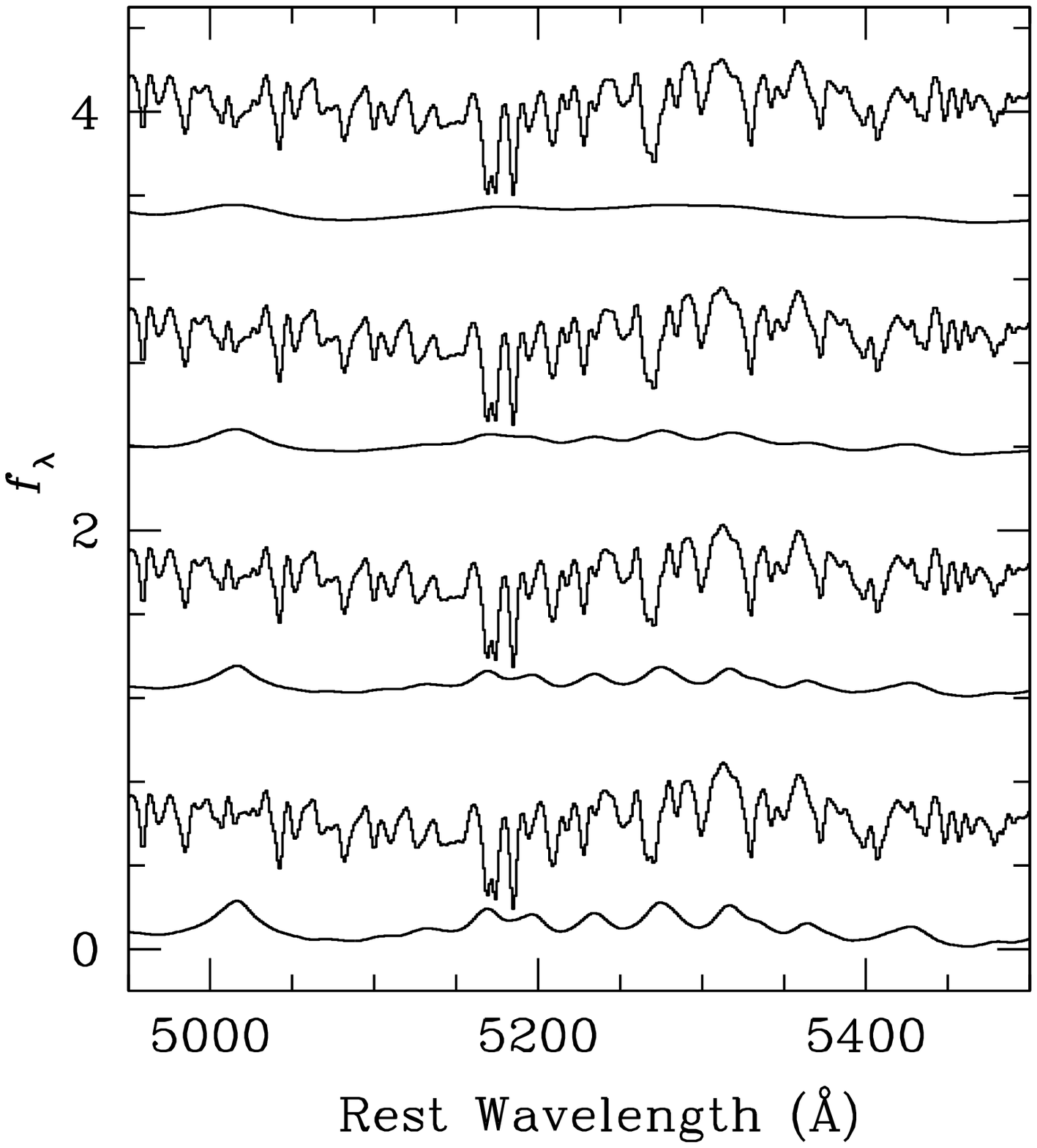,width=10cm,angle=0}
\vskip 1mm 
\figcaption[]{ 
Examples of simulations including Fe~{\tiny
II} contamination.  The stellar continuum is modeled as the average of
10 template stars, while the Fe~{\tiny II} contamination is modeled
using the Fe~{\tiny II} template derived from I~Zw~1.  The template is scaled to Fe~{\tiny II} EWs of 18, 36,
and 54 \AA\ in our models, while the line width is parameterized using
the Eddington ratio (\lledd; see discussion in text).  From top to
bottom we illustrate \lledd\ = 0.01, 0.1, and 1 with EW(Fe~{\tiny II})
= 36 \AA, and \lledd\ = 1 with EW(Fe~{\tiny II}) = 54 \AA.  The bottom
spectrum represents the most extreme contamination examined in our
simulations.  Note that in all cases the effects of the Fe~{\tiny II}
on the stellar features are subtle (for instance, there is a change in
the relative strengths of the Mg~{\tiny I}{\it b} triplet).
\label{fecontamination}}
\vskip 5mm
%%%%%%%%%%%%%%%%%%%%%%%%%%%%%%%%%%%%%%%%%%%%%%%%%%%%%%%%%%%%%%%%%%%
%\noindent

%%%%%%%%%%%%%%%%%%%%%%%%%%%%%%%%%%%%%%%%%%%%%%%%%%%%%%%%%%%%%%%%%%%%
\begin{figure*}
\vbox{ 
\vskip -1.5truein
\hbox{
\psfig{file=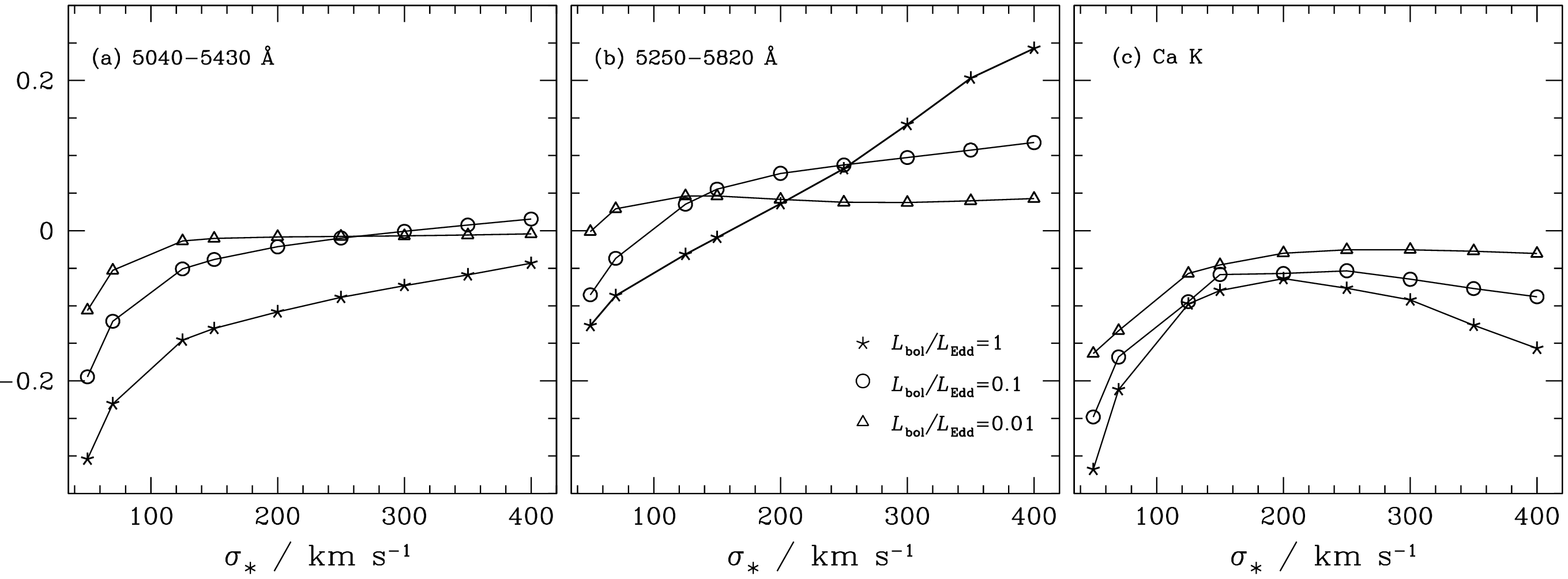,width=0.67\textwidth,angle=-90} 
}}
\vskip -11mm
\figcaption[]{
The effect of Fe~{\tiny II} contamination on \sigmastar.  We compare
the systematic offsets introduced in three different spectral regions:
({\it a}) Mg~{\tiny I}{\it b}, ({\it b}) the Fe region redward of
Mg~{\tiny I}{\it b}, and ({\it c}) Ca~K.  The model spectrum used in
the simulations is the average of 10 template stars, with
Fe~{\tiny II} contamination added using the I~Zw~1 template.
The amplitude of Fe~{\tiny II} is kept constant for
these simulations with an EW of 36 \AA, but its velocity width
decreases as the Eddington ratio increases.  We use the input \sigmastar\
value to calculate \mbh, and the assumed \lledd\ value sets the
continuum luminosity.  We then infer FWHM of \hbeta, which empirically
matches the FWHM of Fe~{\tiny II} (e.g.,~Boroson \& Green 1992). The
percentage systematic error introduced is quantified by \ldelsig\
$\equiv$
[$\sigma{\mathrm{(out)}}-\sigma{\mathrm{(in)}}$]$/\sigma{\mathrm{(in)}}$.
\label{fakefe}}
\end{figure*}
%%%%%%%%%%%%%%%%%%%%%%%%%%%%%%%%%%%%%%%%%%%%%%%%%%%%%%%%%%%%%%%%%%%%
\noindent
\feii\ template.  In so doing we 
introduce three new degrees of freedom: a velocity shift, a velocity
dispersion, and an amplitude for the \feii\ spectrum.  Rather than
allowing these parameters to vary freely, we adopt an iterative
process.  Initially we perform the standard fit as described above
with no \feii\ component.  When \feii\ contamination is
present, the residuals of the initial fit contain significant \feii\
features.  We perform a least-squares fit between the residuals and
the \zw\ template to measure the velocity shift of \feii, using the
region from 5040 to 5820 \AA\ to improve the leverage on the velocity.
The shift is then constrained to this value in the final fit, while
the \feii\ width is fixed to the velocity dispersion of
\hbeta\footnote{We must constrain the Fe {\tiny II} velocity because
large velocity shifts have been observed between the BLR and the
systemic velocity (e.g.,~Marziani \etal\ 1996).  To fit the \hbeta\
velocity dispersion, we first model and remove the narrow \hbeta\
component using a fit constrained by the narrow \halpha\ line fit, as
described in Greene \& Ho (2005b).  The [O {\tiny III}] $\lambda
\lambda$4959, 5007 lines are simultaneously modeled and removed.  We
then fit as many Gaussian components as needed to model the broad
component, and measure the line width from the broad-line model.}
(when we allow this parameter to be free, the code often fails).  In
simulations in which we artificially add \feii\ contamination and then
use this multi-step fitting procedure to remove it, we find that 
we are able to recover the input \sigmastar\ values, with a
slight negative offset of \ldelsig $\approx -5 \%$, for both spectral
regions.  This, however, assumes that we know the width of the \feii\
lines perfectly.  Once we introduce a $20\%$ variation in the \feii\
width, the output dispersions have associated errors of $\sim 10\%$. 

On the other hand, when we attempt to use this fitting procedure with
our actual data, the results are not very encouraging.  A slight
positive systematic difference of 2\%--5\% is introduced between the
fits with and without the \feii\ modeling.  As we have noted, our
objects are not high-\lledd\ objects.  Recall that we selected objects
with both high \cat\ absorption EWs and neglible CaT emission.  Since
CaT emission is correlated with \feii\ emission, which in turn is
correlated with \lledd\ (Persson 1988; Persson \& Ferland 1989;
Boroson 2002), we have predominantly selected AGNs with low \feii\ EW
and low \lledd.  Using virial mass estimates based on \halpha\ line
width and luminosity (Greene \& Ho 2005b), we find
$\langle$log~\lledd$\rangle = -1.3 \pm 0.4$ for this sample (Greene \&
Ho 2005c).  In this regime, when the \feii\ emission is very smooth,
it is somewhat degenerate with the polynomial in the fit.  While the
\feii\ EW estimates from the code range from 1--300 \AA, with a median
value of 80 \AA, we do not believe these values are well-constrained.
It is possible that in more extreme cases, in which the \feii\
emission is more pronounced, it may be possible to obtain a more
robust fit for the \feii\ emission, and thus recover \sigmastar.  Of
course, high-resolution and high-S/N observations would also aid
significantly in performing this procedure.

We conclude that, while broad \feii\ contamination can cause
large biases in \mgb\ measurements, it is not responsible for the
offsets observed in Figure 7{\it a}\ between \mgb\ and \cat.  Instead,
narrow coronal emission is probably the primary culprit, which
suggests that alternate spectral regions with less coronal emission
contamination and, ideally, less \feii\ emission as well, need to be 
explored.  The optical \feii\ multiplets
become negligible at $\sim 5520$~\AA, and the region redward of \mgb\
is rich in metal lines, which, while having lower EWs, in principle
may be used to measure \sigmastar.  We have experimented with the
following spectral regions: 5520--5820 \AA\ and 5520--6280 \AA, which
avoid \feii\ contamination almost completely, as well as both
5040--5820 \AA\ and 5040--6280 \AA.  Finally, we tried a compromise
region, 5250--5820 \AA, which contains the high-EW \ion{Fe}{1}
absorption features, at the expense of some \feii\ contamination.  In
all cases, we mask 5710--5740 \AA\ to remove the
often-present [\ion{Fe}{7}]~$\lambda 5721$ line, and, where relevant,
we mask 5820--5950 \AA\ to exclude the Na~D feature, which is
susceptible to strong interstellar contamination (e.g., Bica et
al. 1991).  In cases including the \mgb\ and \feii\ features we have
experimented with masking the \mgb\ features themselves, as well as
other regions most contaminated by \feii\ emission (specifically
5250--5325 \AA).  We have found that including the spectral region
redward of Na~D introduces significant template mismatch problems, and
therefore do not consider it further.  Of all the possibilities
remaining, we find that the region 5250--5820 \AA\ minimizes the
systematic offset with \cat\ (as shown in Fig. 7{\it b}). With 
$\langle$\ldelsig$\rangle$ = 
$\langle \mathrm{[} \sigma{\mathrm{(Fe)}} -\sigma{\mathrm{(CaT)}}
\mathrm{]} / \sigma{\mathrm{(CaT)}} \rangle$ = $-0.018 \pm$ 0.20, this
region is unambiguously superior to \mgb\ both in terms of systematic
offset and total scatter.  Because narrow coronal emission is common
and introduces a strong systematic uncertainty, we recommend using the Fe
region rather than the \mgb\ region itself.  However, we note that, at
least with the moderate resolution and S/N of SDSS data, it is
preferable to avoid both the Fe and \mgb\ regions when either the AGN
fraction is \gax 85\% or \lledd\ $\approx 1$.

\subsection{Ca H+K}

The \chk\ lines are also high-EW spectral features potentially useful
for obtaining \sigmastar\ measurements.  Traditionally they have been
avoided for a variety of reasons.  Perhaps the most serious
complication is that their line shapes are a strong function of
spectral type, causing systematic errors from template mismatch
(e.g.,~Kormendy \& Illingworth 1982; Kobulnicky \& Gebhardt 2000).
Since H$\epsilon$ overlaps with Ca H and H8 sits directly blueward of
Ca K, young stellar populations may seriously affect the region
(e.g.,~Gebhardt \etal\ 2003).  Apart from line shape, there is a very
steep gradient in the local continuum (the ``4000 \AA\ break'') whose
slope depends on spectral type.  An additional problem is caused by possible
interstellar absorption in the host galaxy, which would lead to narrow
cores (see top panel in Fig. 1).  Finally, the lines have limited
utility at low \sigmastar\ because they are, relatively speaking,
intrinsically broad.  In combination, these factors cause the \chk\
lines to systematically overestimate \sigmastar, as found by both Kormendy \& 
Illingworth (1982) and Bernardi et~al. (2003).  Dressler
(1979), on the other hand, finds no systematic offsets between the
\chk\ region and the G-band (itself sensitive to template mismatch).
However, Dressler was observing a cD galaxy, which, due to the
(presumably exclusively) old stellar population and large velocity
dispersion ($> 375$ \kms) offers the most favorable conditions for
using \chk. For a sample of late-type galaxies, Kobulnicky \& Gebhardt
(2000) obtain reasonable agreement ($\sim 20 \%$) between \chk\ absorption
line widths and kinematics inferred from \ion{H}{1} and \oii\ $\lambda
3727$ emission line widths.  Note, however, that a comparison with
emission features is not the same as a direct comparison with stellar
absorption features.  Also, as emphasized by Kobulnicky \& Gebhardt,
their success hinges on their ability to include a wide range of
spectral types in the modeling.

Of course, AGN contamination only makes the situation more
complicated.  The power-law continuum is rising steeply to the blue,
fractionally increasing the AGN contamination there, and H$\epsilon$ 
(and in principle [\ion{Ne}{3}] $\lambda$3968)
emission renders the Ca H line completely unusable.  On the positive
side, compared to the \mgb\ region, as we saw in \S 4.2, \feii\ contamination
is relatively minimal.  Despite these many complications, \chk\ 
presents two major benefits compared to all other spectral regions.
First of all, it is the {\it only} stellar feature with sufficient EW
to persist in high-luminosity AGNs; for instance, it is detected in
the composite SDSS quasar spectrum of Vanden Berk \etal\ (2001).  Secondly,
the lines are blue enough to remain in the optical bandpass beyond a
redshift of 1.  \chk\ 
represents our only hope of obtaining
\sigmastar\ measurements in powerful AGNs or those at intermediate
redshift\footnote{Of course, there are less direct substitutions for
\sigmastar, such as [O {\tiny III}] line width (e.g.,~Shields \etal\
2003).  While statistically these techniques are accurate, they have
large scatter (see, e.g.,~Greene \& Ho 2005a).}.  Below we investigate
whether, despite its problems, the \chk\ region can provide a useful
diagnostic of \sigmastar\ in AGNs.

%%%%%%%%%%%%%%%%%%%%%%%%%%%%%%%%%%%%%%%%%%%%%%%%%%%%%%%%%%%%%%%%%%%%
\begin{figure*}[t]
\vbox{ 
\vskip -0.9truein
\hbox{
%Bounding Box: 38 15 592 668
\psfig{file=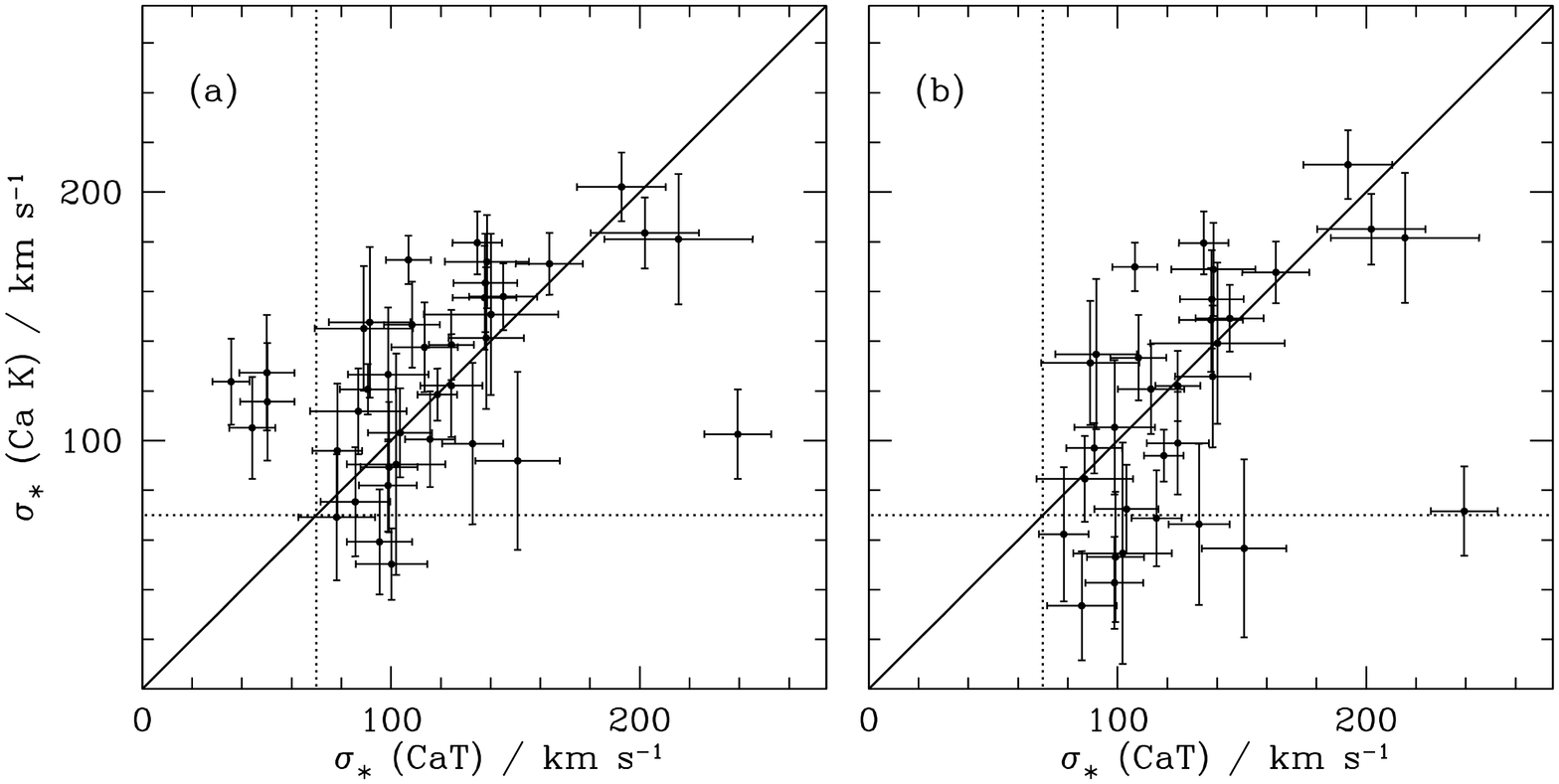,width=0.7\textwidth,keepaspectratio=true,angle=-90}
}}
\vskip -12mm
\figcaption[]{ ({\it a}) A comparison of \sigmastar\ using the \cat\ and Ca~K
spectral regions.  The solid line represents perfect agreement, while
the dotted lines mark the nominal resolution limit of the SDSS.
Error bars are derived as described in \S 3.2.2. ({\it b}) Same as ({\it a}),
except that we have corrected the Ca~K measurements using Equation
2.  It is clear that the correction is too large at low \sigmastar.
\label{cahkcomp}}
\vskip -5mm
\end{figure*}
%%%%%%%%%%%%%%%%%%%%%%%%%%%%%%%%%%%%%%%%%%%%%%%%%%%%%%%%%%%%%%%%%%%

We follow the approach of the previous section, and compare
\sigmastar\ from \chk\ with that from \cat\ to seek the combination of
spectral region, polynomial order, and masking that yields the best agreement.
In terms of the best spectral fitting region, we have found
that broad H$\epsilon$ emission is always a contaminant of Ca H.  Even
when the emission line is not clearly visible, the centroid of the
Ca H line is clearly displaced from its expected position relative to
Ca K, so that the code cannot find a
reasonable solution.  For that reason, we simply mask the Ca H line
(3955--3985 \AA) in all fits; hereafter, our discussion
will focus exclusively on the Ca~K line.  Over this very restricted spectral
region, AGN dilution can be somewhat degenerate with increasing
velocity dispersion.  To mitigate this problem, we include 50 \AA\ of continuum
redward of Ca H in order to help the code determine the general level
of AGN contamination.  Because of the limited spectral region, we also
experimented with fewer polynomial orders or a power-law continuum
constant with wavelength, but in both cases the final \chisq\ values
were elevated, and the fits were unsatisfactory.  Finally, we
attempted a two-step fitting approach to try to mitigate the potential
bias caused by a narrow core from interstellar absorption.  The first
iteration was as described above, while for the second iteration we
masked the line core but simultaneously fixed the velocity shift.
Unfortunately, when only the wings are available, the \sigmastar\
measurements become extremely (unphysically) broad, so this technique
is untenable.  In the end, we settled on a fitting region of 3910--4060 \AA\ 
with Ca H masked.  The results of the comparison with \cat\ are shown in 
Figure 10.  We find $\langle$\ldelsig$\rangle$ = $\langle \mathrm{[} \sigma{\mathrm{(Ca~K)}}-\sigma{\mathrm{(CaT)}} \mathrm{]} / \sigma{\mathrm{(CaT)}}\rangle = 
0.049 \pm 0.29$~\kms, which is fairly reasonable agreement.

As noted above, the \chk\ lines have been known to overestimate
\sigmastar, even in the absence of AGN contamination.  To evaluate the
type of systematic errors that may ensue over a larger range of
\sigmastar\ in the absence of AGN contamination, we measured
$\sigma{\mathrm{(Ca~K)}}$ for the sample of 504 SDSS Type 2 Seyfert
galaxies described in \S3.2.1.  As can be seen in Figure 11,
$\sigma{\mathrm{(Ca~K)}}$ is systematically biased, in a
\sigmastar-dependent way.  At low \sigmastar\ the Ca~K values are too
large, while at large values of \sigmastar\ the Ca~K values are too
small, with the change occurring at \sigmastar\ $\approx 175$ \kms.
Since the comparison for our program objects is between \cat\ and
Ca~K, it may be more appropriate to investigate offsets between these
two spectral regions.  We have therefore compiled an alternate sample
of 374 objects from Heckman \etal\ (2004) that have $z < 0.03$ and a
S/N $> 10$ around the \cat.  Unfortunately, because of the redshift
constraint, the velocity dispersions of the sample are low ($< 225$
\kms), and so we cannot investigate global behavior from these points
alone, but the qualitative offset is similar to that seen in the
large sample (Fig. 12).  This trend is somewhat reminiscent of the
bias apparent in \sigmastar(\chk) shown in Figure 4{\it c}\ due to an A
star component, which suggests that the bias may be due to 

%%%%%%%%%%%%%%%%%%%%%%%%%%%%%%%%%%%%%%%%%%%%%%%%%%%%%%%%%%%%%%%%%%%%
%%BoundingBox: 75 180 570 720
%\hskip -0.3truein
\psfig{file=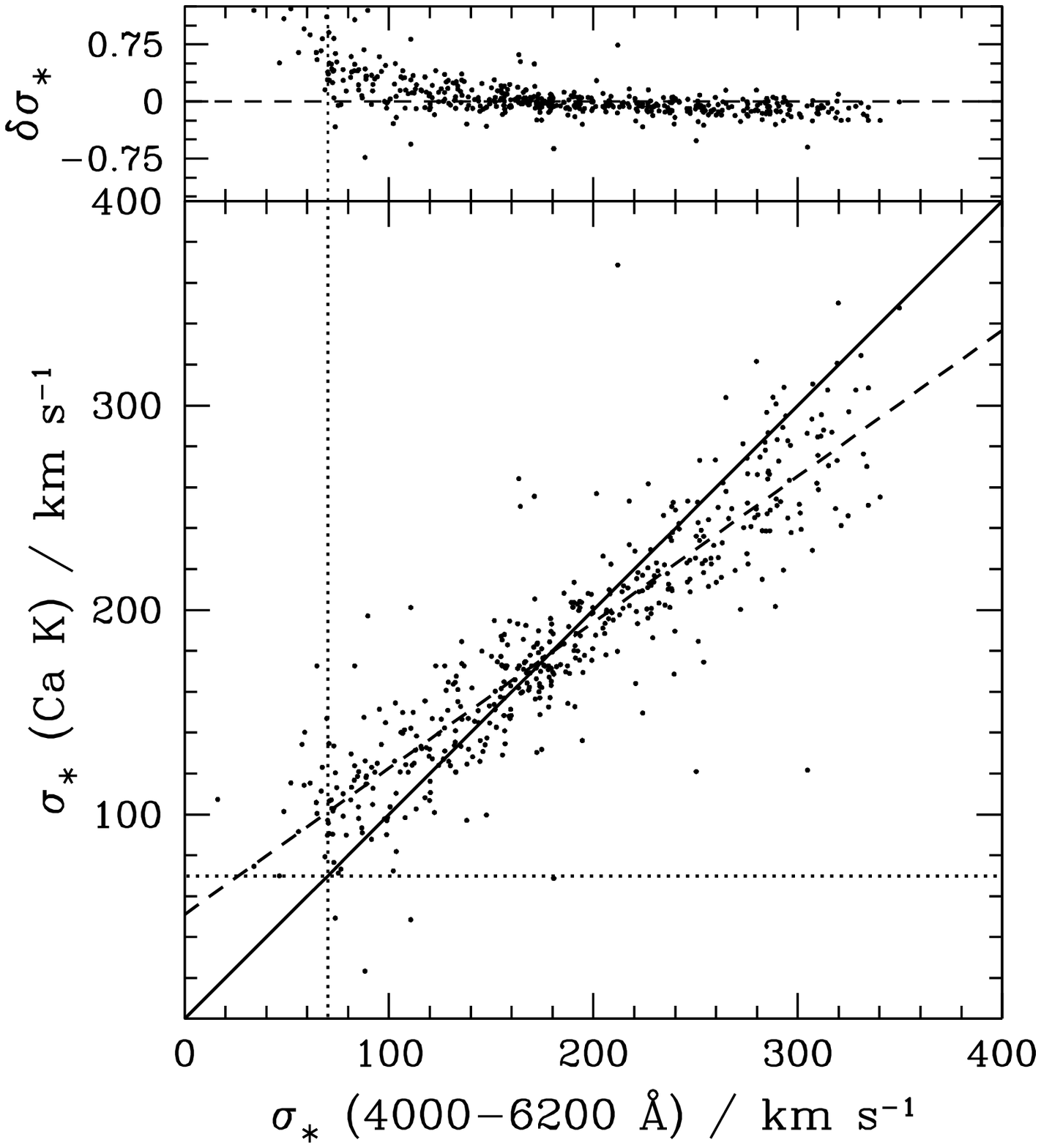,width=0.5\textwidth,keepaspectratio=true,angle=0}
\vskip -2mm 
\figcaption[]{ 
Comparison between \sigmastar\ from
4000--6200 \AA\ as compared to Ca~K, for a sample of 504 Type 2
Seyfert galaxies from Heckman \etal\ (2004).  The solid line
represents equality of the two measurements, and the dotted lines mark
the nominal SDSS resolution limit of 70 \kms.  The dashed line shows
our linear least-squares fit (Equation 2).  The top panel plots the
fractional difference \ldelsig $\equiv$
[$\sigma{\mathrm{(Ca~K)}}-\sigma{\mathrm{(4000-6200
\AA)}}$]$/\sigma{\mathrm{(4000-6200 \AA)}}$.
\label{berncahkcomp}}
\vskip 5mm
%%%%%%%%%%%%%%%%%%%%%%%%%%%%%%%%%%%%%%%%%%%%%%%%%%%%%%%%%%%%%%%%%%%

%%%%%%%%%%%%%%%%%%%%%%%%%%%%%%%%%%%%%%%%%%%%%%%%%%%%%%%%%%%%%%%%%%%%
\begin{figure*}[t]
\vbox{
\vskip 0.1truein
\hbox{
\psfig{file=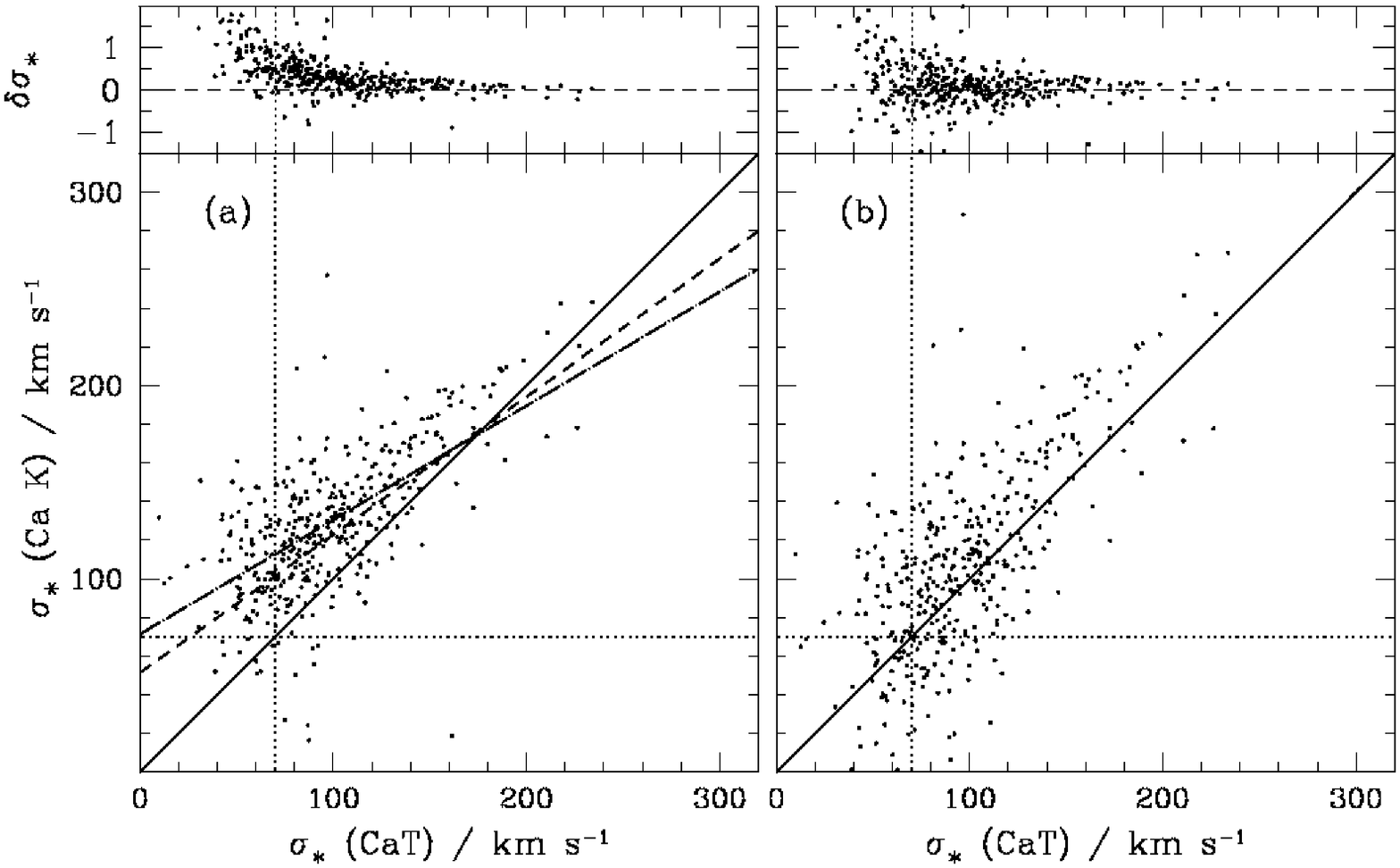,width=0.6\textwidth,keepaspectratio=true,angle=-90}
}}
\vskip -1mm
\figcaption[]{
({\it a}) Velocity dispersion measurements for a sample of 374 Type 2
Seyfert galaxies from Heckman \etal\ (2004), with $z < 0.03$ and S/N
$> 10$ in the \cat\ region.  We have measured $\sigmastar$ using both
the \cat\ and the Ca~K regions.  The solid line denotes
$\sigma$(Ca~K) = $\sigma$(\cat); the dashed line shows the fit derived
from Figure 11 for an alternate set of Seyfert 2 galaxies (Eq. 2; see
text for details), the dash-dot line shows the fit derived for this
sample (Eq. 3), and the dotted lines mark the SDSS resolution limit.
$\langle \mathrm{[}
\sigma{\mathrm{(Ca~K)}}-\sigma{\mathrm{(\cat)}}\mathrm{]}
/\sigma{\mathrm{(\cat)}} \rangle = 0.29 \pm 0.29$~\kms.  ({\it b}) The
same as ({\it a}), except that we have corrected the Ca~K values
using Equation 2, shown in ({\it a}) as a dashed line.  We denote
equality of the two velocity dispersion measures with a solid line.
After the correction, $\langle \mathrm{[}
\sigma{\mathrm{(Ca~K)}}-\sigma{\mathrm{(\cat)}}\mathrm{]}/
\sigma{\mathrm{(\cat)}} \rangle = 0.11 \pm 0.34$~\kms.
\label{cahkcompcorr}}
\end{figure*}
%%%%%%%%%%%%%%%%%%%%%%%%%%%%%%%%%%%%%%%%%%%%%%%%%%%%%%%%%%%%%%%%%%%%

\noindent
differing A star contaminations as a function of \sigmastar.  We
investigate whether young stellar populations are responsible for the
observed offsets by seeking a correlation between \ldelsig\ and the
H$\delta$ index tabulated by Brinchmann \etal\ (2004; Kauffmann \etal\
2003a).\footnote{The H$\delta$ stellar absorption index gives a
measure of the age of the stellar population; younger populations
yield a larger H$\delta$ index.}  In the case of the larger sample the
correlation, while highly significant, is weak (Kendall's $\tau=0.22$,
$P_{\mathrm{null}}=10^{-5}$), while for the \cat\ comparison sample
the correlation strength is even weaker (Kendall's $\tau=0.14$,
$P_{\mathrm{null}}=5 \times 10^{-4}$).  The largest values of
\ldelsig\ correspondingly have the largest H$\delta$ indices, but a
large H$\delta$ index does not ensure a large \ldelsig.  While young
stellar populations apparently do play a role in driving the observed
deviations, they may not fully account for it.  We have also
investigated whether including H$\delta$ as a secondary parameter
improves the correlation between $\sigma$(Ca~K) and $\sigma$(4000-6200
\AA) or $\sigma$(\cat), using the partial Kendall's $\tau$ test from
Akritas \& Siebert (1996).  In both cases the partial correlation
coefficient is lower than the correlation between the two \sigmastar\
variables alone.  We thus do not have a complete explanation for the
offset at this time.

Nevertheless, we can use the trend observed in Figure
11 to derive a correction for the Ca~K velocity dispersions.  We use
an ordinary least-squares fit, and derive the errors using bootstrap
simulations.  We find
\begin{equation}
\sigmastar\ = 1.40 (\pm 0.04)\sigma{\mathrm{(\chk)}} -  71 (\pm 5).
\end{equation}
\noindent
We also calculate a correction using the \cat\ comparison sample only:
\begin{equation}
\sigmastar\ = 1.7 (\pm 0.1)\sigma{\mathrm{(\chk)}} -  121 (\pm 13).
\end{equation}

\noindent
The two corrections are not consistent within the estimated
uncertainties, but when we apply Equation 2 to the \cat\ data set, we
do significantly improve the agreement between $\sigma$(Ca~K) and
$\sigma$(\cat).  Prior to correction, $\langle$\ldelsig$\rangle$ =
$\langle \mathrm{[} \sigma{\mathrm{(Ca~K)}}-\sigma{\mathrm{(\cat)}}
\mathrm{]} /\sigma{\mathrm{(\cat)}} \rangle$ = $0.29 \pm 0.29$.  After
applying the correction from Equation (2; see Fig. 12{\it b}), we find
$\langle$\ldelsig$\rangle$ = $0.11 \pm 0.34$, while when we use the
correction derived from this data set, we find
$\langle$\ldelsig$\rangle = 0.0015\pm0.35$.  Clearly the correction
derived from the \cat\ data is superior in the sense that the mean
difference is closer to zero after the correction is applied.
However, it is worth noting that the scatter increases significantly
($\sim 20\%$) in both cases.  Although the scatter increases, thus
limiting the achievable precision, the most crucial point is that we
can control the {\it systematic}\ bias in the case of the Type 2
objects shown here.

%%%%%%%%%%%%%%%%%%%%%%%%%%%%%%%%%%%%%%%%%%%%%%%%%%%%%%%%%%%%%%%%%%%%
\begin{figure*}
\vbox{
\vskip -0.08truein
\hskip 10mm
\hbox{
\psfig{file=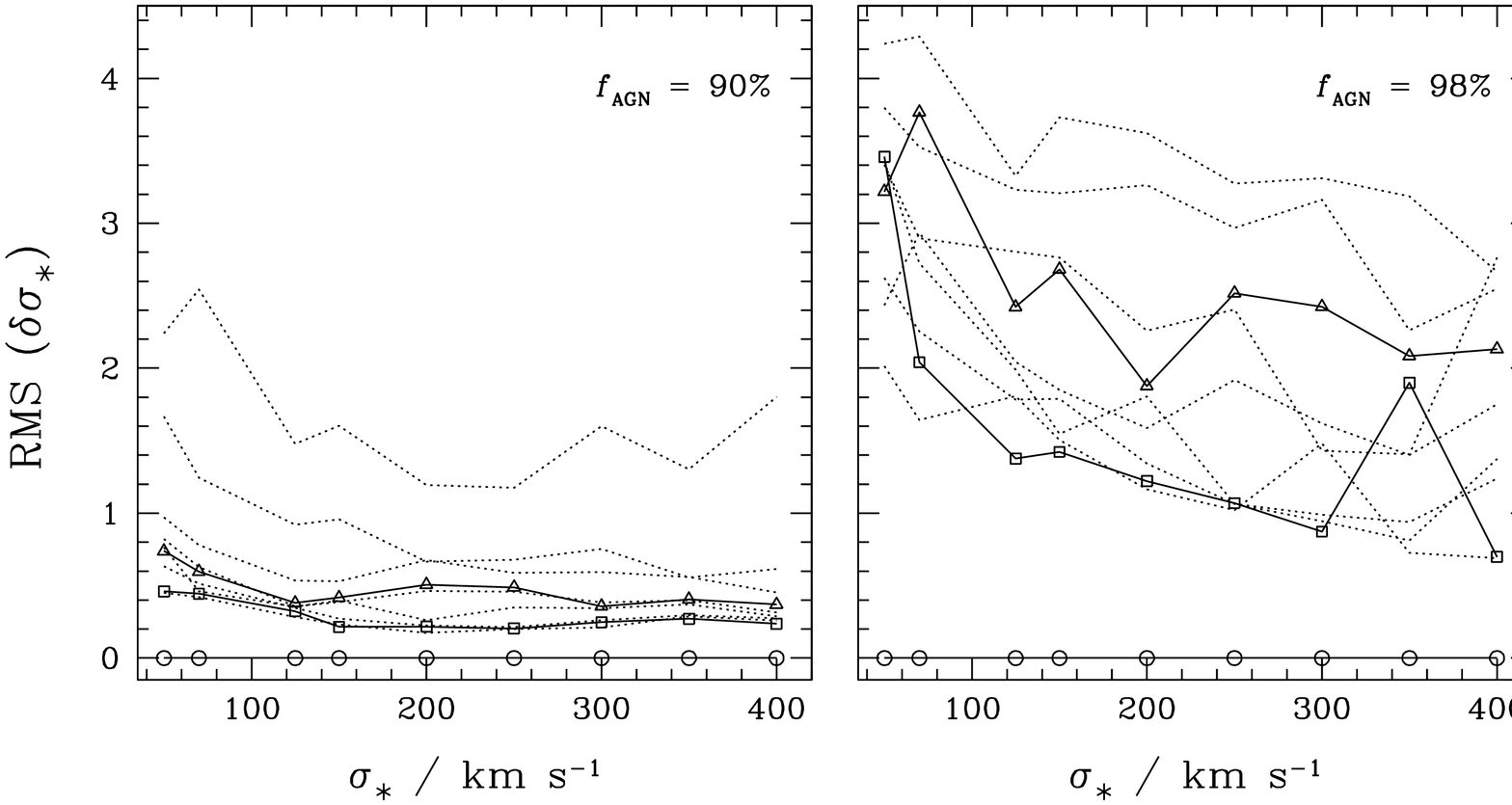,width=0.48\textwidth,angle=-90}
}}
\vskip -2mm
\figcaption[]{
Input-output simulations of
velocity dispersion measurements from the Ca H+K region, similar to
Figure 5, but with higher AGN contamination.
Galaxy models constructed as the average of 10 template stars.
\ldelsig\ $\equiv$
[$\sigma{\mathrm{(out)}}-\sigma{\mathrm{(in)}}$]$/\sigma{\mathrm{(in)}}$.
Simulations were carried out for S/N from 15 to 55, in steps of 5.  
The highest S/N (i.e., no additional noise) is
$\sim$ 95 (shown in {\it open circles}).  As in Figure 5, we highlight in
solid lines S/N = 30 ({\it triangles}) and S/N = 50 ({\it boxes}).
We show an AGN component comprising ({\it left}) 90\% and ({\it right}) 98\% of
the total continuum.
\label{fakecahk}}
\end{figure*}
%%%%%%%%%%%%%%%%%%%%%%%%%%%%%%%%%%%%%%%%%%%%%%%%%%%%%%%%%%%%%%%%%%%%

If this is a valid correction, we should be able to apply it to our
AGNs, a completely independent sample of objects, and improve their
agreement with \cat.  We have performed this test in Figure 10{\it b},
using Equation 2.  In general we are overcorrecting at low \sigmastar,
leading to $\langle$\ldelsig$\rangle$ = 
$-0.15 \pm 0.39$ (the scatter and offset become worse when Eq. 3
is employed).  To determine whether Type 1 AGNs in general can be treated in 
this manner will require a larger database of velocity dispersion measurements
spanning a wider range in \sigmastar.  We are currently assembling just such a
sample (J.~E.~Greene \& L.~C.~Ho, in preparation).  While corrections
of this nature are very dangerous, depending as they do on the
detailed stellar populations of a given sample, we emphasize again
that the \chk\ lines may be the {\it only} available stellar features
for estimating \sigmastar\ in high-luminosity AGNs.  To illustrate
that the Ca~K line can be used even at very high levels of AGN dilution, in
Figure 13 we show simulations for AGN continuum fractions of 90\% and 98\%.
As can be seen, even with an AGN fraction as high as 90\% the Ca~K line
can still provide an estimate of \sigmastar\ to within a factor of 2, for a 
moderate S/N of 20.  However, with 98\% AGN dilution, there is no hope of 
obtaining a meaningful measurement of \sigmastar.

\section{Practical Guidelines}

The optimal spectral region for measuring \sigmastar\
depends on the Eddington ratio and continuum level of the AGN, as well
as the redshift of interest.  For $z < 0.05$, \cat\ is the region of choice.  
In the redshift interval of $0.05 < z < 0.76$, the Fe region
is the best option, provided that the Eddington ratio is \lax 0.5 and
the overall dilution level is \lax 85\%.  Otherwise, the Ca~K line 
must be used.  Finally, for $0.76 < z < 1.3$ Ca~K is the only
available choice (assuming the observations are made in the optical
band).  

We summarize these findings schematically in Figure 14.  For a given
\mbh, we estimate the bulge luminosity using the relation of Marconi
\& Hunt (2003), log (\mbh/\msun) = $1.26 (\pm 0.13$) [log
($L_B/L_{\sun}$)$-10$] $-\,2.96 (\pm 0.48$).  If we make the
simplifying assumption that the entire bulge and only the bulge is
contributing to the starlight, we can then estimate the AGN dilution
in the $B$ band for any AGN luminosity (expressed here in Eddington
units).  Of course, this is an approximation that is
aperture-dependent.  At $z=0.1$, the projected SDSS aperture of 5 kpc
is well matched to the effective radius $R_{\mathrm{eff}} \approx 4$
kpc of a \sigmastar\ = 150 \kms\ bulge (Bernardi \etal\ 2003; Greene
\& Ho 2005a), while at higher $z$ we are more likely underestimating
the total contribution from starlight.  We further caution that the
actual AGN contamination will differ between spectral regions,
becoming worse toward the blue (see \S 4.1).  Finally, there is
significant intrinsic scatter in the \mbh-$L_{\mathrm{bulge}}$
relation; Marconi \& Hunt estimate that the intrinsic scatter is 0.48
dex.  Using corrections from Fukugita \etal\ (1995) to convert from
$m_B$ to $m_i$, we further indicate the maximum redshift to which an
AGN of $m_i=19.1$ mag (the SDSS magnitude limit for AGN-selected
spectroscopic selection) would be detected by the SDSS.  The lines
of constant redshift shift to the right or left for brighter or
fainter apparent magnitudes, respectively.  Bear in mind that these
lines are schematic only.  We do not account for the combined (AGN
plus host galaxy) color of the system, and it would be more
appropriate to apply the SDSS galaxy magnitude limit below a certain
AGN fraction.  However, our goal here is simply to illustrate how the
different constraints define natural regions in the
\mbh-$f_{\mathrm{AGN}}$ plane.

We have assumed that the observations are limited to optical
wavelengths.  Although spectroscopy in the near-infrared is
challenging, due to telluric absorption features and strong and 
variable emission from the night sky, it may be the best option in
some cases.  For instance, narrow-line Seyfert 1 galaxies, thought to
be low-mass BHs at high \lledd\ (e.g.,~Pounds \etal\ 1995; Boller
\etal\ 1996), typically contain such strong \feii\ contamination that
\cat\ may be the only measurable absorption line.  Of course, the
chances of strong \caii\ emission are also highest in these objects,
making the determination of \sigmastar\ particularly challenging
(e.g.,~Botte \etal\ 2004; Barth \etal\ 2005).

\section{Summary}

We present a statistically significant sample of 40 AGNs with
measurable velocity dispersion (\sigmastar) using regions surrounding
the \ion{Ca}{2} triplet, the \mgb\ triplet, and \chk\ lines, for the purposes
of intercomparison.  Using a newly developed direct-fitting code, we
perform a comprehensive set of simulations and 

%%%%%%%%%%%%%%%%%%%%%%%%%%%%%%%%%%%%%%%%%%%%%%%%%%%%%%%%%%%%%%%%%%%%
%%BoundingBox:18 144 592 680
\psfig{file=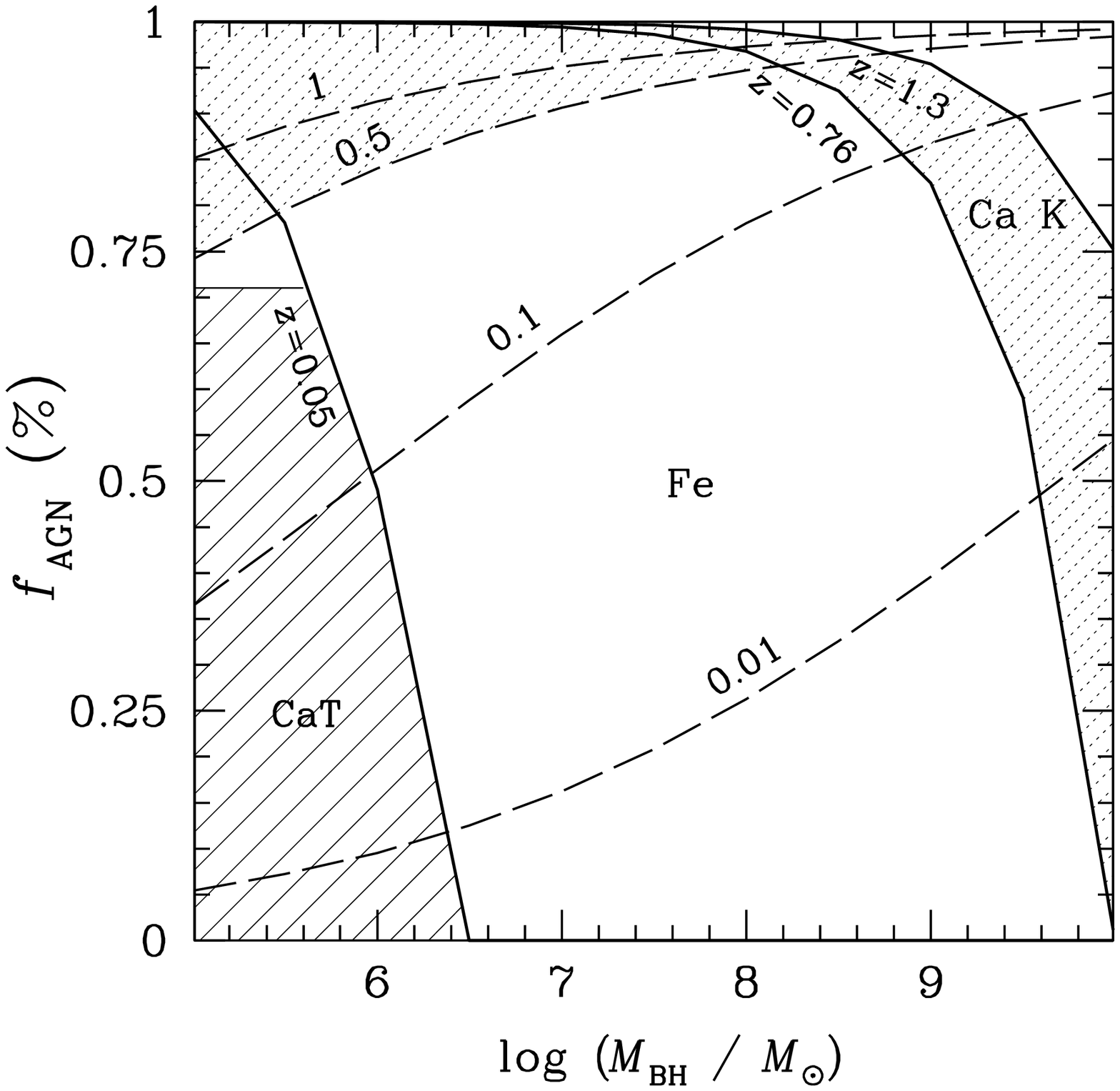,width=0.5\textwidth,keepaspectratio=true,angle=0}
\vskip -5mm
\figcaption[]{
Schematic diagram to delineate the regimes in which the different 
spectral regions can be most effectively used to measure \sigmastar.
The solid curves mark the three key redshift intervals ($z$ = 0.05, 0.76, 
and 1.3) to which an $m_i=19.1$ mag AGN will be spectroscopically
targeted by the SDSS, and the dashed curves identify, from bottom to top, 
\lledd\ = 0.01, 0.1, 0.5, and 1.  The 
fraction of light coming from the bulge was estimated using the
\mbh-$L_{\mathrm{bulge}}$ relation of Marconi \& Hunt (2003).
\label{dil}}
\vskip 5mm
%%%%%%%%%%%%%%%%%%%%%%%%%%%%%%%%%%%%%%%%%%%%%%%%%%%%%%%%%%%%%%%%%%%%
%\noindent
cross-checks to evaluate the merits and limitations of using each spectral
region, with the aim of obtaining realistic estimates of the many
systematic uncertainties that affect measurements of \sigmastar\ in
AGNs.  We argue that the \ion{Ca}{2} triplet is least susceptible to
template mismatch and AGN contamination from emission lines, and so
provides the most reliable measurements of \sigmastar.  We therefore
use these lines as a benchmark to test the other spectral regions.

We examine two types of AGN contamination: featureless continuum
dilution that effectively lowers the S/N of the absorption features,
and emission lines, both narrow and broad, 
that fill in the stellar absorption lines and bias the line profiles
in subtle ways.  In terms of dilution, we find that \sigmastar\ is
measurable for AGN fractions \lax 71\%, \lax 85\%, and \lax 90\% using
\ion{Ca}{2} triplet, \mgb, and \chk, respectively.  As for AGN
emission-line contamination, we find that measurements around \mgb\
are very sensitive both to broad \feii\ emission and narrow emission
from [\ion{Fe}{6}], [\ion{Fe}{7}], and [\ion{N}{1}].  The narrow lines
can be avoided by using the spectral region 5250--5820 \AA.  Broad
\feii\ contamination can bias the fits severely when the \feii\ width
is narrowest.  For a given \mbh\ this will occur at the highest
luminosity, and thus is worst when the object radiates at a large
fraction of its Eddington luminosity.  In such cases, and for the
highest levels of AGN dilution, the \chk\ spectral region (and particularly the
Ca~K line) is the best remaining option.  While there is a systematic
offset between $\sigma$(Ca~K) and \sigmastar\ derived from other
spectral regions, in Type 2 AGNs it is well fit by a linear relation
that allows us to derive unbiased \sigmastar\ estimates from Ca~K.
Further work is required to ensure that this correction is generally
applicable.

As a by-product of this study, we have derived reliable velocity
dispersions for 35 Type 1 AGNs, and set out reasonable selection
criteria to generate a much larger sample of AGNs with a broader range
of $z$, black hole mass, and accretion rates.  Such a sample would
enable us to examine the \msigma\ relation for active galaxies with
true statistical power.  We will be able to calibrate virial
masses for black holes with unprecedented accuracy, as well as
search for second-order trends in the \msigma\ relation with mass,
accretion rate, and redshift.  Greene \& Ho (2005c) present a
preliminary investigation of the \msigma\ relation using the sample
analyzed in this study.

\acknowledgements 
The referee's thorough and insightful comments
significantly improved both the content and the clarity of this manuscript.
We thank A.~J.~Barth for sharing his direct-fitting code, M.~Bernardi
for discussions about her fitting methods, T.~A. Boroson for providing us
with the {\sc I}~Zw~1 iron template, M.~Sarzi for providing his
version of the Rix \& White code, and D.~J.~Schlegel for discussing
his direct-fitting method prior to publication.
L.~C.~H. acknowledges support by the Carnegie Institution of
Washington and by NASA grants from the Space Telescope Science
Institute (operated by AURA, Inc., under NASA contract NAS5-26555).
We are grateful to the SDSS collaboration for providing the
extraordinary database and processing tools that made this work
possible.

%\newpage

\end{document}